\documentclass[letterpaper,english,aps,prl,letterpaper,twocolumn,showpacs,floatfix,groupedaddress,superscriptaddress]{revtex4}
\usepackage{mathptmx}
\usepackage[T1]{fontenc}
\usepackage[latin9]{inputenc}
\usepackage{amsmath}
\usepackage{amssymb}
\usepackage{mathrsfs}
\usepackage{graphicx}
\usepackage{subfigure}

\makeatletter

\newcommand{\noun}[1]{\textsc{#1}}

\makeatother

\usepackage{babel}

\begin{document}

\title{Time-Dependent Density Functional Theory of Open Quantum Systems in the Linear-Response Regime}

\author{David G. Tempel}

\address{Department of Physics, Harvard University,
17 Oxford Street, 02138, Cambridge, MA}

\author{Mark A. Watson}

\author{Roberto Olivares-Amaya}

\author{Al\'an Aspuru-Guzik}

\address{Department of Chemistry and Chemical Biology, Harvard University,
12 Oxford Street, 02138, Cambridge, MA}

\email{aspuru@chemistry.harvard.edu}

\begin{abstract}
Time-Dependent Density Functional Theory (TDDFT) has recently been extended to describe many-body open quantum systems (OQS) evolving under non-unitary dynamics according to a quantum master equation. In the master equation approach, electronic excitation spectra are broadened and shifted due to relaxation and dephasing of the electronic degrees of freedom by the surrounding environment. In this paper, we develop a formulation of TDDFT linear-response theory (LR-TDDFT) for many-body electronic systems evolving under a master equation, yielding broadened excitation spectra. This is done by mapping an interacting open quantum system onto a non-interacting open Kohn-Sham system yielding the correct non-equilibrium density evolution. A pseudo-eigenvalue equation analogous to the Casida equations of usual LR-TDDFT is derived for the Redfield master equation, yielding complex energies and Lamb shifts. As a simple demonstration, we calculate the spectrum of a C$^{2+}$ atom in an optical resonator interacting with a bath of photons. The performance of an adiabatic exchange-correlation kernel is analyzed and a first-order frequency-dependent correction to the bare Kohn-Sham linewidth based on G{\"o}rling-Levy perturbation theory is calculated.

\end{abstract}


\maketitle

\section{I. Introduction}

Due to its attractive balance between accuracy and efficiency, time-dependent density functional theory (TDDFT) has seen a tremendous growth of applications in recent years. These range from optical properties of molecules, clusters and solids, to optimal control theory and real-time dynamics of species in intense laser fields~\cite{runge gross, annu, review of burke, review, rappoport, li 1}.  
	TDDFT has been particularly successful at calculating optical response properties of electronic systems in the linear response regime~\cite{linear response}. In most quantum chemical codes, excitation energies and oscillator strengths are extracted by solving a pseudo-eigenvalue equation, originally formulated by Casida ~\cite{Casida Review}. The Casida equations are derived by considering the linear density response of an interacting system and corresponding non-interacting Kohn-Sham system, both undergoing unitary evolution. If the Casida equations are solved using the ubiquitous adiabatic approximation (ATDDFT) within a discrete basis set, the resulting eigenvalues are real. This gives rise to a discrete absorption spectrum of delta function peaks. 
	
	In experimentally observed spectra, line broadening arises from a variety of different mechanisms, several of which have been explored already within LR-TDDFT. In extended systems, relaxation and dephasing due to electron-electron scattering is well captured using non-adiabatic and current-density dependent functionals~\cite{Vignale-Kohn, VUC, Vignale-Dgosta, Ullrich 2-electron}. In finite systems, decay of resonant states due to coupling with the continuum gives rise to finite linewidths. The ability of DFT and TDDFT to capture lineshape parameters and widths of resonances has been discussed in~\cite{neepa resonance, exx resonance, adam resonance}. 
	
	Another important broadening mechanism arises from relaxation and dephasing of electronic degrees of freedom by a classical or bosonic bath such as photons, phonons or impurities. For extended systems, this situation was considered in~\cite{weakly disordered, ullrich springer,  memory function}, where linewidths of  intersubband plasmon excitations were well captured by combining the Vignale-Kohn functional~\cite{Vignale-Kohn} to account for electron-electron scattering with the memory function formalism for electron-phonon and electron-impurity scattering. For atomic and molecular systems, the theory of open quantum systems within the master equation approach is often used~\cite{nakajima, zwanzig, breuer, Kossakowski72a, lindblad, gorini}. Several important examples include vibrational relaxation of molecules in liquids and solid impurities ~\cite{Mukamel Non-Markov,Mukamel Frank-Condon,  Nitzan} , cavity quantum electrodynamics (QED) ~\cite{resonator, spontaneous, spontaneous1, spontaneous2}, photo-absorption of chromophores in a protein bath~\cite{Redfield Absorption, Redfield Absorption 2, protien dynamics, mukamel 1}, single-molecule transport~\cite{burke, gebauer chemphyschem, gebauer prl, chen 1, chen 2} and exciton transport~\cite{exciton transfer, nano letter, ENAQT, mukamel 2}.  In all of these examples, even with simple system-bath models, the exact solution of the master equation for the reduced dynamics of the many-body electronic system is computationally intractable. Therefore, open quantum systems TDDFT (OQS-TDDFT) offers an attractive approach to the many-body open-systems problem.
	
	Several important steps toward the formulation of OQS-TDDFT have recently been made, with the focus on real-time dynamics. In~\cite{burke}, a Runge-Gross theorem was established for Markovian master equations of the Lindblad form. A scheme in which the many-body master equation is mapped onto a non-interacting Kohn-Sham master equation was proposed for application to single-molecule transport. In~\cite{joel's paper, prl}, the Runge-Gross theorem was extended to arbitrary non-Markovian master equations and a Van Leeuwen construction was established, thereby proving the existence of an OQS-TDDFT Kohn-Sham scheme~\cite{van leeuwen}. In~\cite{prl}, it was shown that the original interacting open system dynamics can be mapped onto either a non-interacting open Kohn-Sham system, or a non-interacting closed (unitarily evolving)  Kohn-Sham system. A different formulation of OQS-TDDFT based on the stochastic Schrodinger equation has also been developed~\cite{d prl, da prb, d prb}.
	
	The goal of the present manuscript is to formulate the linear response version of (OQS-TDDFT) within the master equation approach. This provides a framework in which environmentally broadened spectra of many-body electronic systems can be accessed in an \textit{ab initio} way using TDDFT, especially when combined with microscopically derived master equations. We use the scheme discussed in~\cite{prl, burke}, in which the interacting OQS can be mapped onto a non-interacting open Kohn-Sham system, yielding the same density response. This scheme is better suited to response theory than the closed Kohn-Sham scheme also discussed in~\cite{prl}, since relaxation and dephasing is already accounted for in the Kohn-Sham system. The unknown (OQS-TDDFT) exchange-correlation functional only needs to correct the relaxation and dephasing in the Kohn-Sham system to that of the interacting system, rather than needing to explicitly account for the entire effect of the environment. However,  the closed Kohn-Sham scheme is better suited for real-time dynamics, since one only needs to propagate a set of equations for the Kohn-Sham orbitals as in usual TDDFT. This is in contrast to the open Kohn-Sham scheme, in which $N^2-1$ equations are propagated for the elements of the density matrix, with N being the dimensionality of the Hilbert space.
	
	The paper is organized as follows.  In Section II,  we formulate the most general OQS-TDDFT linear response equations for arbitrary non-Markovian master equations with initial correlations. In section III, we make the treatment more specific by focusing on the Redfield master equation. We also derive Casida-type equations whose solution yields the environmentally broadened absorption spectrum. The solutions to these equations are complex, with the real part of the frequency yielding the location of absorption peaks and the imaginary part yielding the linewidths. In Section IV, we apply the formalism developed in Section III to a C$^{2+}$ atom in an optical resonator setup evolving under the Redfield master equation. Section V analyzes the performance of using an adiabatic functional (OQS-ATDDFT) in solving the OQS-TDDFT Casida equations derived in section III. To a large degree, OQS-ATDDFT is seen to provide a reliable correction to the location of absorption peaks while leaving the linewidths unchanged. A frequency-dependent functional yielding a correction to the OQS-ATDDFT linewidth based on G{\"o}rling-Levy (GL) perturbation theory is then calculated and analyzed. In section VI a discussion and outlook is provided. 
	
Atomic units in which $e = \hbar = m_{e} = 1$ are used throughout. This also implies that the speed of light in vacuum is given by $c= 137$. For generality, we have formulated most of the theory by considering linear response from an equilibrium state at finite temperature. For atoms and molecules, it will generally be sufficient to take the zero temperature limit and consider linear response from the ground-state.




\section{II. General formulation of OQS-TDDFT linear response theory}

\subsection*{A. Linear response of interacting open quantum systems}

Our formulation of the interacting OQS density-density response function parallels that used in~\cite{spin response} for calculating spin susceptibilities (see also~\cite{Linear Response Open, reduced response}). The starting point is the unitary evolution for the full density matrix of the system and the reservoir (we use the terms "reservoir" and "bath" interchangeably throughout) ,

\begin{equation}
\frac{d}{dt} \hat{\rho}(t) = \frac{1}{\imath} [\hat{H}(t), \hat{\rho}(t)]  \equiv - \imath \breve{\mathscr{L}}(t)  \hat{\rho}(t),
\end{equation}

where $\breve{\mathscr{L}}(t)$ is the Liouvillian superoperator for the full system and reservoir dynamics. The full Hamiltonian is given by

\begin{equation}
\hat{H}(t) = \hat{H}_{S}(t) + \hat{H}_{R} + \hat{V}.
\end{equation}

Here,

\begin{equation}
H_{S}(t) = -\frac{1}{2} \sum_{i=1}^N \nabla_{i}^2 + \sum_{i<j}^{N} \frac{1}{|\mathbf{r}_i - \mathbf{r}_j|} + \sum_i v_{ext}(\mathbf{r}_i, t),
\label{many-body hamiltonian}
\end{equation}

is the Hamiltonian of the electronic system of interest in an external potential $v_{ext}(\mathbf{r}, t)$. This potential generally consists of a static external potential due to the nuclei and an external driving field coupled to the system such as a laser field.  The system-bath coupling, $\hat{V}$, is at this point arbitrary, but we will discuss specific forms later. $\hat{V}$ acts in the combined Hilbert space of the system and reservoir and so it couples the two subsystems. $\hat{H}_{R}$ is the Hamiltonian of the reservoir, assumed to have a dense spectrum of eigenstates. The density of states of $\hat{H}_{R}$ determines the structure of reservoir correlation functions, whose time-scale in turn determines the reduced system dynamics.

Defining the reduced density operator for the electronic system alone by tracing over the reservoir degrees of freedom, 

\begin{equation}
\hat{\rho}_S(t) = Tr_R\{ \hat{\rho}(t)  \},
\end{equation}

one arrives at the formally exact quantum master equation,

\begin{equation}
\frac{d}{dt} \hat{\rho_{S}}(t) = -\imath [\hat{H}_{S}(t), \hat{\rho_{S}}(t)] + \int_{t_{0}}^{t} d\tau  \breve{\Xi}(t-\tau) \hat{\rho_{S}}(\tau) + \Psi(t).
\label{non-markovian}
\end{equation}

Here, $\breve{\Xi}(t-\tau)$ is the memory kernel and $\Psi(t)$ arises from initial correlations between the system and its environment. It is referred to as the inhomogeneous term. The above equation is still formally exact, as $ \hat{\rho_{S}}(t)$ gives the exact expectation value of any observable depending only on the electronic degrees of freedom. In practice, however, approximations to $\breve{\Xi}$ and $\Psi$ are required. Of particular importance in TDDFT is the time-dependent electronic density,

\begin{equation}
n(\mathbf{r}, t) = Tr_S \{ \hat{\rho_{S}}(t) \hat{n}(\mathbf{r}) \},
\end{equation}

 where $\hat{n}(\mathbf{r}) = \sum_i^{N} \delta(\mathbf{r} - \hat{\mathbf{r}}_i)$. We now assume that for $t<t_0$, the external potential is time-independent while for $t>t_0$ a weak perturbing field is applied. i.e.

\begin{eqnarray}
&t& < t_0, v_{ext}(\mathbf{r}, t) = v_{ext}(\mathbf{r})
\\ & t& > t_0, v_{ext}(\mathbf{r}, t) = v_{ext}(\mathbf{r}) + \delta v_{ext}(\mathbf{r}, t).
\end{eqnarray}

For $t<t_0$, the entire system and environment is in thermal equilibrium described by the canonical density operator

\begin{equation}
\hat{\rho}^{eq} = \frac{e^{- \beta \hat{H}}}{Tr_{S+R} \{e^{- \beta \hat{H}} \}},
\label{full equilibrium density matrix}
\end{equation}

where $\beta = \frac{1}{K_B T}$ is the inverse temperature. The reduced equilibrium density operator of the electronic system is then given by

\begin{equation}
\hat{\rho_S}^{eq} =  \frac{Tr_R \{ e^{- \beta \hat{H}} \}}{Tr_{S+R} \{e^{- \beta \hat{H}} \}} .
\label{electron equilibrium density matrix}
\end{equation}

In Eq.~\ref{full equilibrium density matrix} and Eq.~\ref{electron equilibrium density matrix},  $\hat{H} = \hat{H}_{S}+ \hat{H}_{R} + \hat{V}$ is the full Hamiltonian for $t<t_0$ and $\hat{H}_{S} =  -\frac{1}{2} \sum_{i=1}^N \nabla_{i}^2 + \sum_{i<j}^{N} \frac{1}{|\mathbf{r}_i - \mathbf{r}_j|} + \sum_i v_{ext}(\mathbf{r}_i)$ is the static Hamiltonian of the electrons in the absence of the external perturbation. 

For $t>t_0$, the perturbing field is switched on and the system density operator subsequently evolves under the master equation given in Eq.~\ref{non-markovian}. The electronic density evolution to first-order in the perturbing field is then given by

\begin{eqnarray}
&t& < t_0, n(\mathbf{r}, t) = n^{eq}(\mathbf{r})
\\ & t& > t_0, n(\mathbf{r}, t) = n^{eq}(\mathbf{r}) + \delta n(\mathbf{r}, t).
\label{density response first}
\end{eqnarray}

Here, $n^{eq}(\mathbf{r}) = Tr_S \{ \hat{\rho}_S^{eq} \hat{n}(\mathbf{r}) \} $ is the equilibrium electron density and 

\begin{equation}
\delta n(\mathbf{r}, t) = \int d^3\mathbf{r'} \int dt' \chi_{nn}(\mathbf{r}, t; \mathbf{r'}, t') \delta v_{ext}(\mathbf{r'}, t')
\end{equation}

is the linear density response. $\chi_{nn}(\mathbf{r}, t; \mathbf{r'}, t')$ is the density-density response function. Its Fourier transform to the frequency domain is given by

\begin{equation}
\chi_{nn}(\mathbf{r}, \mathbf{r'}; \omega) = \lim_ {\epsilon \rightarrow +0} \imath \int_{0}^{\infty} dt e^{- \imath \omega t - \epsilon t} Tr_{S+R}\{ [\hat{n}(\mathbf{r}, t), \hat{n}(\mathbf{r'})] \hat{\rho}^{eq} \},
\label{non-markovian response transform}
\end{equation}

where 

\begin{equation}
\hat{n}(\mathbf{r}, t) = e^{\imath \hat{H} t} \hat{n}(\mathbf{r}) e^{-\imath \hat{H} t}
\end{equation}

is the operator generating the electronic charge density in the Heisenberg picture with respect to the full Hamiltonian for $t<t_0$. Rearranging terms under the trace operation in Eq.~\ref{non-markovian response transform}, the density-density response function can be written as

\begin{equation}
\chi_{nn}(\mathbf{r}, \mathbf{r'}; \omega) = \lim_ {\epsilon \rightarrow +0} \imath \int_{0}^{\infty} dt e^{- \imath \omega t - \epsilon t} Tr_{S} \{ \hat{n}(\mathbf{r}) \hat{\rho}_{S}^{n}(\mathbf{r'}, t) \}.
\label{non-markovian response reduced}
\end{equation}

Here, $\hat{\rho}_{S}^{n}(\mathbf{r}, t)$ is an operator acting in the electronic system Hilbert space, which obeys the same equation of motion as the reduced system density operator (Eq.~\ref{non-markovian})

\begin{equation}
\frac{d}{dt} \hat{\rho}_{S}^{n}(\mathbf{r}, t) = -\imath [\hat{H}_{S}, \hat{\rho}_{S}^{n}(\mathbf{r}, t)] + \int_{t_{0}}^{t} d\tau  \breve{\Xi}(t-\tau) \hat{\rho}_{S}^{n}(\mathbf{r}, \tau) + \Psi(t),
\label{non-markovian dipole}
\end{equation}

subject to the initial condition

\begin{equation}
 \hat{\rho}_{S}^{n}(\mathbf{r}, 0) = Tr_R\big\{ [\hat{n}(\mathbf{r}), \hat{\rho}^{eq}] \big\}.
\end{equation}

Carrying out the Fourier transform in Eq.~\ref{non-markovian response reduced}, one arrives at the formally exact expression for the open-systems density-density response function in Liouville space,

\begin{equation}
\chi_{nn}(\mathbf{r}, \mathbf{r'}; \omega) = \imath Tr_S \Bigg\{ \hat{n}(\mathbf{r}) \frac{1}{ \omega + \breve{\mathscr{L}}_{S} - \imath \breve{\Xi}(\omega)} (\hat{\rho}_{S}^{n}(\mathbf{r'}, 0) + \Psi(\omega))\Bigg\}.
\label{non-markovian response}
\end{equation}

$\breve{\mathscr{L}}_{S}$ is the Liouvillian for the system Hamiltonian for $t<t_0$, defined by it's action on an arbitrary operator $\hat{O}$ by $\breve{\mathscr{L}}_S  \hat{O} = [\hat{H}_S, \hat{O}]$. It is readily verified that Eq.~\ref{non-markovian response} reduces to the usual expression for the density-density response function of an isolated system when $\breve{\Xi}(\omega) =0$, $\Psi(\omega) = 0$ and $\hat{\rho}_S^{n}(\mathbf{r}, 0) = [\hat{n}(\mathbf{r}), \hat{\rho}_S^{eq}]$. 

The absorption spectrum can be extracted by taking the imaginary part of $\chi_{nn}(\mathbf{r}, \mathbf{r'}; \omega)$ in Eq.~\ref{non-markovian response}. For an isolated system with a discrete spectrum, this is given by a sum over weighted delta function peaks. For an open system as in Eq.~\ref{non-markovian response}, $\breve{\Xi}(\omega)$ and $\Psi(\omega)$ in principle give rise to the exact complicated broadened and shifted spectrum, due to relaxation and dephasing of the electronic degrees of freedom by the environment. In practice, however, even with simple approximations to $\breve{\Xi}(\omega)$ and $\Psi(\omega)$, the exact form of $\Im m \left[\chi_{nn}(\mathbf{r}, \mathbf{r'}, \omega)\right]$ is not exactly known, since it refers to a many-body response function. In the next two subsections, we consider an open Kohn-Sham system, formally yielding the exact density response of the original interacting open system. In an open-systems TDDFT framework, the exact spectrum of Eq.~\ref{non-markovian response} is obtained by correcting the open Kohn-Sham spectrum via an exchange-correlation kernel. The kernel must take into account not only the electron-electron interaction contained in $\breve{\mathscr{L}}_{S}$, but must also correct the interaction of the system with the bath, described by $\breve{\Xi}(\omega)$ and $\Psi(\omega)$.

\subsection*{B. The open Kohn-Sham system}

It was proven in~\cite{prl}, that for a master equation of the form given in Eq.~\ref{non-markovian}, there exists a unique, non-interacting and open Kohn-Sham system, whose system density operator evolves under the master equation

\begin{equation}
\frac{d}{dt} \hat{\rho}_{S}^{ks}(t) = -\imath [\hat{H}^{ks}(t), \hat{\rho}_{S}^{ks}(t)] + \int_{t_{0}}^{t} d\tau  \breve{\Xi}^{ks}(t-\tau) \hat{\rho}_S^{ks}(\tau) + \Psi^{ks}(t),
\label{KS non-markovian}
\end{equation}

such that the time-dependent density is obtained from

\begin{equation}
n(\mathbf{r}, t)  = Tr \{ \hat{\rho}_S^{ks}(t) \hat{n}(\mathbf{r}) \} 
\label{exact density}
\end{equation}

for all times. $\hat{H}^{ks}(t) = \sum_{i=1}^{N} \hat{h}_i^{ks}(t)$, where the Kohn-Sham Hamiltonian is given by 

\begin{equation}
h^{ks}(\mathbf{r}, t) = -\frac{1}{2} \nabla^2 + v_{ks}(\mathbf{r}, t). 
\end{equation}

Here, $v_{ks}$ is a local, multiplicative, one-body potential which drives the open Kohn-Sham system in such a way that the true density of the original interacting open system is reproduced for all times. In analogy to usual TDDFT, the Kohn-Sham potential is partitioned as 

\begin{equation}
v_{ks}(\mathbf{r}, t) = v_{ext}(\mathbf{r}, t) + v_{h}(\mathbf{r}, t) + v_{xc}^{open}(\mathbf{r}, t), 
\end{equation}

where $v_{h}(\mathbf{r}, t)$ is the Hartree potential and the unknown functional $v^{open}_{xc}(\mathbf{r}, t)$ accounts for electron-electron interaction within the system as well as interaction between the system and bath. In general,

\begin{equation}
v_{xc}^{open}(\mathbf{r}, t) = v_{xc}^{open}(\mathbf{r}, t)[n, \breve{\Xi}, \breve{\Xi}^{ks}, \Psi, \Psi^{ks}, \hat{\rho}_S(0), \hat{\rho}^{ks}(0)].
\label{functional dependence}
\end{equation}

Formally, the open-systems exchange-correlation potential is a functional not only of the density, but also of the memory kernel, inhomogeneous term and initial state of both the interacting and Kohn-Sham systems. It has been shown in usual TDDFT of closed systems, that initial state dependence can be absorbed as dependence on the history of the density and vice versa~\cite{neepa history}. Interestingly, in the theory of open quantum systems, it is possible to absorb the inhomogeneous term $\Psi$, into the memory kernel $\breve{\Xi}$~\cite{tannor}. This raises the possibility that  $v_{xc}^{open}$ may be a functional only of n, $\breve{\Xi}$ and $\breve{\Xi}^{ks}$, but a more rigorous study of this will be done in future work. For notational convenience, we suppress the explicit functional dependence of $v_{xc}^{open}$ on these quantities, although it is implied unless otherwise stated.

 In general, $\breve{\Xi}^{ks}$ and $\Psi^{ks}$ can be chosen to simplify $v^{open}_{xc}(\mathbf{r}, t)$ as much as possible, although with some restrictions \cite{prl} and consistency conditions between $\breve{\Xi}^{ks}$ and $\Psi^{ks}$ \cite{tannor}.

If the system is started in an equilibrium state, as is typically the case in linear response theory, the initial state dependence in Eq.~\ref{functional dependence} is automatically removed.  The equilibrium density, $n^{eq}(\mathbf{r})$, is obtained by solving the Kohn-Sham-Mermin equations~\cite{kohn-sham}

\begin{equation}
\left[ -\frac{1}{2} \nabla^2 + v_{ks}^{eq}[n](\mathbf{r}) \right] \phi_{i}(\mathbf{r}) = \epsilon_{i} \phi_{i}(\mathbf{r}).
\label{mermin equation}
\end{equation}

The Kohn-Sham-Mermin potential is partitioned as

\begin{equation}
v_{ks}^{eq}[n](\mathbf{r}) = v_{ext}[n](\mathbf{r}) + v_{h}[n](\mathbf{r}) + \frac{\delta F_{xc}[n]}{\delta n(\mathbf{r})},
\label{mermin potential}
\end{equation}

where $F_{xc}[n]$ is the exchange-correlation contribution to the free energy. After solving Eq.~\ref{mermin equation}, the equilibrium Kohn-Sham-Mermin density operator is obtained by populating the orbitals according to

\begin{equation}
\hat{\gamma}_{eq}^{ks} = \sum_{i=1}^{\infty} f_{i} |\phi_i \rangle \langle \phi_i|,
\label{mermin density matrix}
\end{equation}

where $f_i$ are Fermi-Dirac occupation numbers

\begin{equation}
f_{i} = \frac{1}{e^{\beta(\epsilon_i -\mu)}+1}.
\end{equation}

Denoting $\langle \mathbf{r}|\phi_i \rangle =\phi_{i}(\mathbf{r})$, the one-particle Kohn-Sham-Mermin density matrix is

\begin{equation}
\langle \mathbf{r}|\hat{\gamma}_{eq}^{ks}| \mathbf{r'} \rangle \equiv  \gamma(\mathbf{r}, \mathbf{r'}) = \sum_{i=1}^{\infty} f_{i} \phi_i^*(\mathbf{r}) \phi_i(\mathbf{r'}).
\end{equation}

The equilibrium density is then obtained by taking the diagonal elements in real space,

\begin{equation}
n^{eq}(\mathbf{r}) = \gamma(\mathbf{r}, \mathbf{r})= \sum_{i=1}^{\infty} f_i |\phi_{i}(\mathbf{r})|^2.
\end{equation}

For the open Kohn-Sham scheme to be useful practically, $v^{open}_{xc}[n](\mathbf{r}, t)$, $\breve{\Xi}^{ks}$ and  $\Psi^{ks}$ should be constructed so that the following conditions are satisfied:

1)  $\breve{\Xi}^{ks}$ and  $\Psi^{ks}$  should not induce correlations between non-interacting electrons as the Kohn-Sham system evolves. This ensures that the N-body Kohn-Sham density matrix in Eq.~\ref{KS non-markovian} can be traced over N-1 electron coordinates to arrive at a closed equation of motion for the Kohn-Sham reduced 1-particle density matrix. Physically, this is expected since most reasonable bath models couple to the electronic system through one-body operators~\cite{Cohen, breuer}. 

 2) The equation of motion for the non-equilibrium open Kohn-Sham reduced 1-particle density matrix should have the Kohn-Sham-Mermin density matrix as its stationary-point solution. This ensures that the system thermalizes to the correct equilibrium density. This also means that at equilibrium, $v^{open}_{xc}[n](\mathbf{r}, t)$ should reduce to $\frac{\delta F_{xc}[n]}{\delta n(\mathbf{r})}$. Although this is automatically satisfied by using $v_{ks}^{eq}[n]$ as an adiabatic approximation, it might not be satisfied by more sophisticated approximations with memory dependence~\cite{neepa history, neepa memory, neepa memory perturbations, baer memory}.


\subsection*{C. Linear response of the open Kohn-Sham system}

Returning to linear response, for $t<t_0$ the Kohn-Sham system is in thermal equilibrium with its environment at inverse temperature $\beta$, described by Eq.~\ref{mermin equation}.  At $t=t_0$, the perturbation $\delta v_{ext}(\mathbf{r}, t)$ is switched on and the Kohn-Sham system subsequently evolves according to Eq.~\ref{KS non-markovian}. The Kohn-Sham potential is given by

\begin{eqnarray}
&t& < t_0, v_{ks}(\mathbf{r}, t) = v_{ks}^{eq}[n](\mathbf{r})
\\ & t& > t_0, v_{ks}(\mathbf{r}, t) = v_{ks}^{eq}[n](\mathbf{r}) + \delta v_{ks}(\mathbf{r}, t),
\end{eqnarray}

where $\delta v_{ks}(\mathbf{r}, t) = \delta v_{ext}(\mathbf{r}, t) + \delta v_{h}[n](\mathbf{r}) + \delta v^{open}_{xc}[n](\mathbf{r}, t)$. Due to Eq. (\ref{exact density}), the exact linear density response of Eq. (\ref{density response first}) is obtained through

\begin{equation}
\delta n(\mathbf{r}, t) = \int d^3\mathbf{r'} \int dt' \chi_{nn}^{ks}(\mathbf{r}, t; \mathbf{r'}, t') \delta v_{ks}(\mathbf{r'}, t').
\end{equation}

Here, $\chi_{nn}^{ks}(\mathbf{r}, t; \mathbf{r'}, t')$ is the density-density response function of the open Kohn-Sham system. It's Fourier transform to the frequency domain is given by

\begin{eqnarray}
&& \chi_{nn}^{ks}(\omega, \mathbf{r}, \mathbf{r'}) =  \imath Tr \Bigg\{ \hat{n} (\mathbf{r})\nonumber \\ &\times&  \frac{1}{ \omega + \breve{\mathscr{L}}_{ks} - \imath \breve{\Xi}^{ks}(\omega)}  ([ \hat{n}(\mathbf{r'}), \hat{\rho}_S^{ks}(0)]+ \Psi^{ks}(\omega))\Bigg\},
\label{Kohn-Sham non-markovian response}
\end{eqnarray}

where $\breve{\mathscr{L}}_{ks}$ is the Liouvillian for the equilibrium Kohn-Sham-Mermin Hamiltonian. Since the system is in the equilibrium state at $t=0$, $\hat{\rho}_S^{ks}(0)$ must yield the equilibrium density, implying that it reduces to the Kohn-Sham-Mermin density matrix when traced over N-1 electron coordinates. We now define the open-systems exchange-correlation kernel in analogy to usual TDDFT for closed-systems by,

\begin{equation}
f_{xc}^{open}[n^{eq}](\mathbf{r}, \mathbf{r'}; \omega) = \frac{\delta v^{open}_{xc}[n](\mathbf{r}, \omega)}{\delta n(\mathbf{r'}, \omega)}|_{n = n^{eq}},
\end{equation}

which is a functional of the equilibrium density. As in TDDFT for closed systems, the interacting and Kohn-Sham response functions are related through a Dyson-like equation,

\begin{eqnarray}
&& \chi_{nn}(\omega, \mathbf{r}, \mathbf{r'}) = \chi_{nn}^{ks}(\omega, \mathbf{r}, \mathbf{r'}) \nonumber \\ &+& \int d^3 \mathbf{y} d^3 \mathbf{y'} \chi_{nn}^{ks}(\omega, \mathbf{r}, \mathbf{y})  \Big\{\frac{1}{|\mathbf{y} - \mathbf{y'}|} + f_{xc}^{open}[n^{eq}](\mathbf{y}, \mathbf{y'}; \omega)\Big\}  \nonumber \\ &\times& \chi_{nn}(\omega, \mathbf{y'}, \mathbf{r'}).
\label{open dyson}
\end{eqnarray} 

$\chi_{nn}^{ks}$, being much simpler then the original interacting $\chi_{nn}$, can readily be constructed from the orbitals and eigenvalues in Eq.~\ref{mermin equation} and approximations to $\breve{\Xi}^{ks}(\omega)$ and $\Psi^{ks}(\omega)$ in terms of these quantities. This will be done explicitly for the Redfield master equation in the next section. Since correlation between the open Kohn-Sham system and reservoir is already partially captured through $\breve{\Xi}^{ks}(\omega)$ and $\Psi^{ks}(\omega)$, the bare Kohn-Sham absorption spectrum extracted from $\Im m \left[\chi_{nn}^{ks}(\omega, \mathbf{r}, \mathbf{r'}) \right]$ is already broadened and shifted. The functional $f_{xc}^{open}[n^{eq}]$ has the task of correcting the spectrum extracted from $\chi_{nn}^{ks}$ to that of the interacting $\chi_{nn}$, incorporating both the usual electron-electron correlation in closed-systems TDDFT as well as additional system-bath correlation. In general, the memory kernel $\breve{\Xi}(\omega)$ may give rise to a very complicated non-analytic structure of $\chi_{nn}$ in the lower half of the complex plane. However, for Markovian master equations, it will be seen that the pole structure of $\chi_{nn}$ in the discrete part of the spectrum consists of simple poles in the lower complex plane, shifted by a finite amount off of the real axis. In such cases,  it might be reasonable to approximate $f_{xc}^{open}$ as

\begin{equation}
f_{xc}^{open}[n^{eq}](\mathbf{r}, \mathbf{r'}; \omega) = \frac{\delta^2 F_{xc}[n]}{\delta n(\mathbf{r})\delta n(\mathbf{r'})}|_{n=n^{eq}} + f_{xc}^{bath}[n^{eq}](\mathbf{r}, \mathbf{r'}; \omega).
\label{fxc partition}
\end{equation}

Here, the first term is just the adiabatic contribution to the exchange-correlation kernel and the second term is an in general frequency-dependent and imaginary correction. This is attractive, since we can take advantage of the usual good performance of adiabatic TDDFT in describing the location of absorption peaks, and attempt to build functionals that go beyond the adiabatic approximation to account for line broadening and lamb shifts. This strategy will be discussed further in section V.

\section{III. LR-TDDFT for the redfield master equation.}

In section II, we formulated LR-TDDFT for a very general class of master equations. In this section, we make the discussion more specific by invoking the Markov approximation and second Born approximation in the system-bath interaction, to arrive at the microscopically-derived Redfield master equation~\cite{Redfield Absorption, Cohen, May/Kuhn, van Kampen, redfield original 2, redfield original}. Since the Redfield equations are rigorously obtained without phenomenological parameters, they are amenable to an~\textit{ab initio} theory such as TDDFT. Although we focus on Redfield theory here, the generalization of our formalism to other Markovian master equations can be made with small modifications. 
Finally, we discuss how it is possible to extract the absorption spectrum of a many-body system evolving under the Redfield equations directly within OQS-TDDFT. This is done by formulating Casida-type equations yielding complex eigenvalues due to coupling with the bath.


\subsection*{A. The Markov approximation and the Redfield master equation.}

The Markov approximation describes a situation in which the bath correlation functions decay on an infinitely fast time-scale relative to the thermalization time of the system~\cite{May/Kuhn, Cohen}.  As a result, the bath has no memory  and the memory kernel is time-local

\begin{equation}
\breve{\Xi}(t-\tau) \propto \breve{\mathscr{R}} \delta(t - \tau).
\end{equation}

Additionally, this implies that the initial density operator is a tensor product of a density operator in the system space with the equilibrium density operator of the bath

\begin{equation}
\hat{\rho}(0)  = \hat{\rho}_s(0) \otimes \Bigg\{ \frac{e^{-\beta \hat{H}_R}}{Tr_{R}\{ e^{-\beta \hat{H}_R}\}} \Bigg\}.
\label{uncorrelated initial}
\end{equation}

As a result of Eq.~\ref{uncorrelated initial}, the system and environment have no initial correlations, and

\begin{equation}
\Psi(t) = 0.
\end{equation}

The master equation (Eq.~\ref{non-markovian}) then takes the simple form,

\begin{equation}
\frac{d}{dt} \hat{\rho}_S (t) = - \imath \breve{\mathscr{L}}_{S}(t) \hat{\rho}_S(t) + \breve{\mathscr{R}} \hat{\rho}_S(t).
\label{markovian master equation first}
\end{equation}

If the system Hamiltonian is time-independent, Eq.~\ref{markovian master equation first} is written in a basis of eigenstates of $\hat{H}_S$ as:

\begin{equation}
\frac{d}{dt} \rho_{ab}(t) = -\imath \omega_{ab} \rho_{ab}(t)+ \sum_{abcd} R_{abcd} \rho_{cd}(t).
\end{equation}

Here, $ \omega_{ab} = E_a - E_b$ are many-body transition frequencies of $\hat{H}_S$ and $R_{abcd}$ are matrix elements of $\breve{\mathscr{R}}$ in this basis. So far our discussion applies to any Markovian master equation. To obtain the Redfield equations, we further assume that the system-bath coupling has a bilinear form

\begin{equation}
\hat{V} = -\hat{S} \otimes \hat{R},
\label{bilinear}
\end{equation}

where $\hat{R}$ is an operator in the reservoir Hilbert space which couples to a local one-body operator $\hat{S} = \left[ \sum_{i=1}^N \hat{S}(\hat{\mathbf{p}}_i, \hat{\mathbf{r}}_i) \right]$ in the system Hilbert space. This form of the system-bath coupling is very general and can apply to electron-phonon coupling in molecules and solid impurities, but also momentum dependent couplings which are relevant for instance in laser cooling, brownian motion in liquids or dissipative strong field dynamics~\cite{tannor, Kohen, Cohen}. The Redfield tensor is then derived by performing second-order perturbation theory in $\hat{V}$, and is given explicitly by

\begin{eqnarray}
&& R_{abcd} = - \int_{0}^{\infty} d \tau \left[g(\tau)\left[\delta_{bd} \sum_{n} S_{an}S_{nc} e^{\imath \omega_{cn} \tau} - S_{ac} S_{db} e^{\imath \omega_{ca} \tau}  \right] \right] \nonumber \\ &-& \int_{0}^{\infty} d \tau \left[g(-\tau)\left[\delta_{ac} \sum_{n} S_{dn}S_{nb} e^{\imath \omega_{nd} \tau} - S_{ac} S_{db} e^{\imath \omega_{bd} \tau}  \right] \right].
\label{interacting redfield}
\end{eqnarray}

For a detailed derivation of the Redfield equations see~\cite{May/Kuhn}. In Eq.~\ref{interacting redfield}, 

\begin{equation}
S_{ab} = N \int d^3 \mathbf{r} \int d^3 \mathbf{r}_2 ... d^3 \mathbf{r}_N \psi^*_{a}(\mathbf{r}, \mathbf{r}_2, ...\mathbf{r}_N) S(\frac{\mathbf{\nabla}}{\imath}, \mathbf{r}) \psi_{b}(\mathbf{r}, \mathbf{r}_2, ...\mathbf{r}_N) 
\end{equation}

are matrix elements of $\hat{S}(\hat{\mathbf{p}}, \hat{\mathbf{r}})$ between system many-body wavefunctions and 

\begin{equation}
\omega_{ab} = E_a - E_b
\end{equation}

are system many-body excitation energies. $g(\tau)$ are bath correlation functions given by

\begin{equation}
g(\tau) = Tr_R \{ \hat{R}(\tau) \hat{R}(0) \},
\end{equation}

where 

\begin{equation}
\hat{R}(\tau) = e^{\imath \hat{H}_R \tau} \hat{R} e^{-\imath \hat{H}_R \tau}.
\end{equation}

\subsection*{B. Linear response of a many-body system evolving under the Redfield master equation.}

Since we consider linear response from the equilibrium state, the initial density matrix for the system is given by

\begin{equation}
\hat{\rho}_{S}(0) = \frac{e^{-\beta \hat{H}_S}}{Tr_S\{ e^{-\beta \hat{H}_S} \} },
\end{equation}

and the density-density response function in Eq.~\ref{non-markovian response} reduces to

\begin{equation}
\chi_{nn}(\mathbf{r}, \mathbf{r'}; \omega) = \imath Tr_S \Bigg\{ \hat{n}(\mathbf{r}) \frac{1}{ \omega + \breve{\mathscr{L}}_{s} - \imath \breve{\mathscr{R}}} [\hat{n}(\mathbf{r'}), \hat{\rho}_{S}(0)] \Bigg\}.
\label{response}
\end{equation}

Inserting a complete set of eigenstates of $\hat{H}_S$ in Eq.~\ref{response}, a sum over states expression for the density-density response function is given by,

\begin{eqnarray}
&& \chi_{nn}(\mathbf{r}, \mathbf{r'}; \omega) \nonumber \\ &=&   \imath \sum_{a} [P(E_a)] \sum_b \Bigg\{ \frac{\langle a| \hat{n}(\mathbf{r})| b \rangle \langle b|\hat{n}(\mathbf{r'})|a \rangle }{\omega +\omega_{ab}+\imath R_{abab}} \nonumber \\ &-& \frac{\langle a| \hat{n}(\mathbf{r'})| b \rangle \langle b|\hat{n}(\mathbf{r})|a \rangle }{\omega -\omega_{ab}+\imath R_{abab}^*} \Bigg\}.
\label{redfield response}
\end{eqnarray}
Here, $P(E_a) = \frac{e^{-\beta E_a}}{\sum_b e^{-\beta E_b} }$ are equilibrium occupation probabilities of the various many-body states. 

By hermiticity of the density matrix, it can be readily verified that $R_{abab}^* = R_{baba}$. We can separate the real and imaginary parts of $R_{abab}$ as

\begin{equation}
R_{abab} = \Gamma_{ab} + \imath \Delta_{ab}.
\label{redfield complex}
\end{equation}

From the pole structure of Eq.~\ref{redfield response}, we see that $\Gamma_{ab}$ corresponds to an imaginary part of the energy of the transition $\omega_{ab}$, giving rise to a finite lifetime, while $\Delta_{ab}$ is a Lamb shift of the real part of the energy. The effect of the Redfield tensor is to shift the poles of the density-density response function by a finite amount into the lower half of the complex plane.







\subsection*{C. The Markovian Kohn-Sham-Redfield equations}


We now discuss the properties of the open Kohn-Sham system for the Redfield master equation. As discussed in section II B, there is some freedom in the construction of the Kohn-Sham dissipative superoperator and corresponding exchange-correlation potential. In this section, we choose a very natural form for the Kohn-Sham superoperator, whose construction is discussed below.

 We consider an open Kohn-Sham system evolving under a Markovian master equation

\begin{equation}
\frac{d}{dt} \hat{\rho}_S^{ks}(t) = - \imath [\hat{H}^{ks}(t), \hat{\rho}_S^{ks}(t)] + \breve{\mathscr{R}}^{ks} \hat{\rho}_S^{ks}(t),
\label{ks markovian master equation}
\end{equation}

which reproduces the exact density evolution of the interacting Markovian master equation in Eq.~\ref{markovian master equation first}. To satisfy condition 1 in section II B, we must have

\begin{equation}
\breve{\mathscr{R}}^{ks}(\mathbf{r}_1,  \mathbf{r}_2, ... \mathbf{r}_N) \equiv \sum_{i=1}^N \breve{r}^{ks}(\mathbf{r}_i),
\label{sum}
\end{equation}

i.e. the N-body Kohn-Sham dissipative superoperator is a sum of one-body superoperators acting on each coordinate separately. This also follows very naturally from the assumed one-body nature of the system-bath interaction in Eq.~\ref{bilinear}. Since $\hat{H}^{ks}(t) = \sum_{i=1}^{N} \hat{h}^{ks}_i(t)$ is also a sum of one body terms,  we can trace both sides of Eq.~\ref{ks markovian master equation} over N-1 electron coordinates and arrive at a closed equation of motion for the Kohn-Sham 1-particle reduced density matrix,

\begin{equation}
\frac{d}{dt} \hat{\gamma}(t) = - \imath [\hat{h}^{ks}(t), \hat{\gamma}(t)] + \breve{r}^{ks} \hat{\gamma}(t).
\label{one -particle ks markovian master equation}
\end{equation}

We can now write Eq.~\ref{one -particle ks markovian master equation} in a basis of Kohn-Sham-Mermin orbitals as

\begin{equation}
\frac{d}{dt} \gamma_{ij}(t) = -\imath \sum_k \Big\{ h^{ks}(t)_{ik} \gamma_{kj}(t) - \gamma_{ik}(t) h^{ks}(t)_{kj} \Big\} + \sum_{ijkl} r^{ks}_{ijkl} \gamma_{kl}(t).
\label{kohn-sham redfield}
\end{equation}

We choose $r^{ks}_{ijkl}$ to have the form of the Redfield tensor, but written in terms of Kohn-Sham-Mermin orbitals and eigenvalues,

\begin{eqnarray}
&& r_{ijkl}^{ks} = - \int_{0}^{\infty} d \tau \left[g(\tau)\left[\delta_{jl} \sum_{m} S_{im}S_{mk} e^{\imath \omega_{km}^{ks} \tau} - S_{ik} S_{lj} e^{\imath \omega_{ki}^{ks} \tau}  \right] \right] \nonumber \\ &-& \int_{0}^{\infty} d \tau \left[g(-\tau)\left[\delta_{ik} \sum_{m} S_{lm}S_{mj} e^{\imath \omega_{ml}^{ks} \tau} - S_{ik} S_{lj} e^{\imath \omega_{jl}^{ks} \tau}  \right] \right]. 
\label{kohn-sham tensor}
\end{eqnarray}

Here,

\begin{equation}
\omega_{ij}^{ks} = \epsilon_i - \epsilon_j
\end{equation}

are bare Kohn-Sham-Mermin transition frequencies and 

\begin{equation}
S_{ij} = \int d^3 \mathbf{r} \phi_{i}^*(\mathbf{r}) S(\frac{\mathbf{\nabla}}{\imath}, \mathbf{r}) \phi_{j}(\mathbf{r})
\end{equation}

are matrix elements of the system-bath coupling operator between Kohn-Sham-Mermin orbitals. 

Eq.~\ref{kohn-sham redfield} has a number of desirable properties. First, it's stationary point solution is the Kohn-Sham-Mermin density matrix, Eq.~\ref{mermin density matrix}, if $\hat{h}^{ks}(t)$ reduces to the Kohn-Sham-Mermin Hamiltonian when evaluated on the equilibrium density. This ensures that condition 2 in section II B is satisfied. It also satisfies detailed balance as well as most other properties of the usual many-body Redfield equations, but in terms of Kohn-Sham-Mermin quantities. Also, the tensor $r^{ks}_{ijkl}$ has a simple form and can be constructed explicitly in terms of orbitals and eigenvalues obtained in an equilibrium-state Kohn-Sham calculation. The potential $v_{xc}^{open}(t)$ contained in $\hat{h}^{ks}(t)$ will in general be a functional of $\breve{r}^{ks}$ and $ \breve{\mathscr{R}}$ as well as the time evolving density.



\subsection*{D. Linear response of the Kohn-Sham-Redfield system and the open-systems Casida equations}

We now consider the open-systems LR-TDDFT formalism developed in section II, but applied to the Redfield master equation. The density-density response function of the Kohn-Sham-Redfield system is given by

\begin{equation}
\chi_{nn}^{ks}(\mathbf{r}, \mathbf{r'}; \omega) = \imath Tr_S \Bigg\{ \hat{n}(\mathbf{r}) \frac{1}{ \omega + \breve{\mathscr{L}}_{ks} - \imath \breve{\mathscr{R}}^{ks}} [\hat{n}(\mathbf{r'}), \hat{\rho}_S^{ks}(0)] \Bigg\}.
\label{markovian response}
\end{equation}

Using Eqs.~53-56 and inserting a complete set of Kohn-Sham-Mermin states, one obtaines the sum-over-states expression,

\begin{eqnarray}
&& \chi_{nn}^{ks}(\mathbf{r}, \mathbf{r'}, \omega) =  \sum_{i}  f_i \sum_j \Bigg\{ \frac{\langle i| \hat{n}(\mathbf{r})| j \rangle \langle j|\hat{n}(\mathbf{r'})|i \rangle  }{\omega +\omega^{ks}_{ij}+\imath r_{ijij}^{ks}}\nonumber \\  & -& \frac{\langle i| \hat{n}(\mathbf{r'})| j \rangle \langle j|\hat{n}(\mathbf{r})|i \rangle }{\omega -\omega^{ks}_{ij}+\imath {r_{ijij}^{ks}}^*} \Bigg\}.
\label{sos markovian response}
\end{eqnarray}

Eq.~\ref{redfield response} and Eq.~\ref{sos markovian response} are related through the Dyson-like relation given in Eq.~\ref{open dyson}. To extract the poles of the interacting density-density response function in Eq.~\ref{redfield response} from that of the Kohn-Sham system in Eq.~\ref{sos markovian response}, a pseudo-eigenvalue equation must be solved for the squares of the complex transition frequencies,

\begin{equation}
\Big\{\omega^2 - \bar{\Omega}(\omega) \Big\} \vec{F} = 0.
\label{open casida}
\end{equation}

The operator $\bar{\Omega}(\omega)$ can be written as a matrix in a basis of Kohn-Sham molecular orbitals (assuming a closed shell system) as

\begin{eqnarray}
&& \bar{\Omega}_{ijkl}(\omega) = \delta_{ik} \delta_{jl} \Big\{ (\omega_{lk}^{ks} + \Delta_{kl}^{ks})^2 + (\Gamma_{kl}^{ks})^2 - 2 \imath \omega \Gamma_{kl}^{ks} \Big\} +\nonumber \\ && 4 \sqrt{(f_i - f_j)(\omega_{ji}^{ks} + \Delta_{ij}^{ks})} K_{ijkl}(\omega) \sqrt{(f_k - f_l)(\omega_{lk}^{ks} + \Delta_{kl}^{ks})}.
\label{mo casida}
\end{eqnarray}

The explicit derivation of Eq.~\ref{open casida} and Eq.~\ref{mo casida} is given in appendix A. In Eq.~\ref{mo casida}, 

\begin{eqnarray}
&& K_{ijkl}(\omega) = \int d^3 \mathbf{r} \int d^3 \mathbf{r'} \phi_i^*(\mathbf{r}) \phi_{j}^*(\mathbf{r}) \Bigg\{ \frac{1}{|\mathbf{r} - \mathbf{r'}|}  \nonumber \\ && +f_{xc}^{open}[n^{eq}, \breve{\mathscr{R}},\breve{r}^{ks} ](\mathbf{r}, \mathbf{r'}; \omega) \Bigg\} \phi_k(\mathbf{r'}) \phi_{l}(\mathbf{r'})
\label{coupling matrix}
\end{eqnarray}

and the bare Kohn-Sham linewidths and Lamb shifts are given by the relation

\begin{equation}
r_{ijij}^{ks} = \Gamma^{ks}_{ij} + \imath \Delta^{ks}_{ij},
\end{equation}

as in Eq~\ref{redfield complex}. In principle, with the exact functional $f_{xc}^{open}[n^{eq}, \breve{\mathscr{R}},\breve{r}^{ks} ]$, the exact poles of Eq.~\ref{redfield response} are recovered by solving Eq.~\ref{open casida}. {The operator $\bar{\Omega}(\omega)$ is non-Hermitian, giving rise to complex eigenvalues corresponding to broadened excitation spectra. $\bar{\Omega}(\omega)$ is also frequency-dependent and imaginary, both explicitly through the third term in Eq.~\ref{mo casida} and implicitly through $f_{xc}^{open}$ in the coupling matrix $K_{ijkl}(\omega)$. The explicit frequency-dependence arises because the bare Kohn-Sham transitions are already broadened, even in the absence of Hartree-exchange-correlation effects. This is most easily seen by setting $K_{ijkl}(\omega) = 0$ in Eq.~\ref{mo casida}. $\bar{\Omega}_{ijkl}$ is then diagonal and Eq.~\ref{open casida} reduces to a set of uncoupled equations given by

\begin{equation}
\Big\{ (\omega_{ij}^{ks} + \Delta_{ij}^{ks})^2 + (\Gamma_{ij}^{ks})^2 - 2 \imath \omega \Gamma_{ij}^{ks} \Big\} F_{ij} = \omega^2 F_{ij}.
\end{equation}

These are solved with the quadratic formula to yield 

\begin{equation}
\omega = - \imath \Gamma_{ij}^{ks} \pm (\omega_{ij}^{ks} + \Delta_{ij}^{ks}),
\end{equation}

which are precisely the poles of Eq.~\ref{sos markovian response}.

\section{IV. Application - Spectrum of a C$^{2+}$ atom from the Redfield master equation}

As a simple demonstration, in this section we calculate the absorption spectrum of an atom in an optical resonator using the Redfield master equation~\cite{Cohen, breuer}. The modes of the radiation field act as a bosonic bath, leading to decay of the atomic excitations due to spontaneous and stimulated emission~\cite{resonator,spontaneous1, spontaneous2, spontaneous}.  We consider a hypothetical experimental setup similar to that used in~\cite{resonator}. The effect of the optical resonator is to modify the density of radiation field modes relative to the vacuum, leading to an enhancement of the decay rate of atomic excitations at certain frequencies~\cite{resonator, Cohen, quantum optics review}. Although we do not have access to experimental data for C$^{2+}$ in an optical resonator cavity, we use accurate experimental data taken in vacuum, together with a chosen cavity geometry to construct a numerically exact spectrum~\cite{NIST}.


\subsection*{Spontaneous emission from the Redfield master equation}

For a single atom with zero center of mass velocity contained in an optical resonator cavity, the system-bath interaction is  


\begin{equation}
\hat{V} = -\imath \vec{\hat{\mu}} \cdot \sum_{i} \vec{\epsilon}_i \sqrt{\frac{\omega_{i}}{2 V}} (a_i - a_i^{\dag}).
\end{equation}

Here, $\vec{\hat{\mu}}$ is the dipole operator for the atom, $\vec{\epsilon}_i$ and $\omega_i$ are respectively the polarization vector and frequency of the ith mode of the radiation field and V is the quantization volume. $a_i$ and $a_i^{\dag}$ respectively destroy and create a photon in the ith mode of the cavity. The photon reservoir Hamiltonian is

\begin{equation}
\hat{H}_R = \sum_{i} \omega_{i} a_i^{\dag} a_i.
\end{equation}

With the system-bath interaction and bath hamiltonian specified, the Redfield tensor can be explicitly constructed~\cite{Cohen}. 

The entire atom-field system is taken to be in thermal equilibrium at inverse temperature $\beta$, such that $\omega_{01} \gg \frac{1}{\beta}$. With this condition, the atom can be assumed to be in it's groundstate and the effect of stimulated emission is neglected. We then need to only construct the matrix elements $R_{a0a0}$ appearing in Eq.~\ref{redfield response}. For the real part of $R_{a0a0}$ one finds

\begin{equation}
\Gamma_{a0} = \frac{(2 \pi)^2}{V} \int \int |\vec{\mu}_{a0} \cdot \vec{\epsilon}_i |^2 \omega_k g(\omega_k, \vec{k}) \delta(\omega_{a0} - \omega_k) d\Omega_k d \omega_k.
\label{linewidth}
\end{equation}

The imaginary part of $R_{0a0a}$ is given by 

\begin{equation}
\Delta_{a0} = \frac{2 \pi}{V} \sum_{n} \mathscr{P} \int \int |\vec{\mu}_{an} \cdot \vec{\epsilon}_i |^2 \frac{\omega_k}{\omega_{an} - \omega_k} g(\omega_k, \vec{k}) d\Omega_k d \omega_k,
\label{lamb shift}
\end{equation}

where $\mathscr{P}$ denotes the principle value integral~\cite{Cohen}. $\omega_k$ is the frequency of a photon whose wave-vector magnitude is $k=|\vec{k}|$. $\omega_{an} = E_a - E_n$ is the difference in atomic energy levels and $\vec{\mu}_{an}$ are matrix elements of the dipole operator between atomic wavefunctions. $g(\omega)$ is the density of field modes in the cavity.  In free-space, the density of field modes takes the form  $g_{free}(\omega) = \frac{V \omega^2}{(2 \pi)^2 c^3}$. This gives rise to a free-space natural linewidth

\begin{equation}
\Gamma_{a0}^{free} = \frac{4  \mu_{a0}^2 \omega_{a0}^3}{3 c^3}.
\end{equation}

If one considers an experimental setup such as that used in~\cite{resonator}, the density of field modes is modified to

\begin{equation}
\rho(\omega) = g_{free}(\omega) M(\omega),
\end{equation} 

where the function $M(\omega)$ is given by

\begin{equation}
M(\omega) = \frac{\sqrt{1+F}}{1+F {\rm sin}^2(\frac{\omega L}{c})}.
\label{modification}
\end{equation}

The linewidth in the cavity is then modified to

\begin{equation}
\Gamma_{a0} = \frac{4  \mu_{a0}^2 \omega_{a0}^3}{3 c^3}M(\omega_{a0}),
\label{cavity linewidth}
\end{equation}

where we have neglected cavity edge effects and angular dependence of $\rho$. In Eq.~\ref{modification}, L is the length of the cavity and $F = \frac{4 R}{(1-R)^2}$, where R is the reflection coefficient of the cavity walls. By changing the mirror reflectivity and cavity length, suppression or enhancement of the spontaneous emission rate is possible. For our calculations, we choose cavity parameters of $R=0.998$ and $L=4.88$ cm, leading to an overall enhancement. Using experimental data for the atomic energy levels and the transition dipole matrix elements of C$^{2+}$ taken from~\cite{NIST}, together with the specified parameters for the cavity geometry, we can explicitly construct Eq. ~\ref{cavity linewidth}. We include in our calculation the 3 lowest dipole allowed transitions of C$^{2+}$, which are $1s^2 2s^2 \rightarrow 1s^22s (2p, 3p, 4p)$. The numerical values of the linewidths (imaginary part of the frequency) calculated in Eq.~\ref{cavity linewidth} are given in the fourth column of Table~\ref{table: imaginary part}. Due to lack of experimental data on all atomic transitions, the Lamb shifts in Eq.~\ref{lamb shift} cannot be explicitly evaluated. They are, however, estimated to be several orders of magnitude smaller and will be neglected in the following analysis. From the transition dipole matrix elements we can also construct the oscillator strengths. The "exact" response function constructed with these parameters is included in Figures 1-3. 

\subsection*{OQS-ATDDFT calculation of the spectrum of C$^{2+}$}

As a first step, we solve Eq.~\ref{open casida} using only an adiabatic approximation to the exchange-correlation kernel. This corresponds to including the first term in Eq.~\ref{fxc partition}, but entirely neglecting $ f_{xc}^{bath}(\mathbf{r}, \mathbf{r'}; \omega)$. 

For our calculations, we obtain the Kohn-Sham parameters of C$^{2+}$ to be inputted in Eq.~\ref{mo casida} using the real-space TDDFT package Octopus~\cite{octopus 1, octopus 2, octopus 3}. First, a ground-state DFT calculation is performed using the local density approximation (LDA) with the modified Perdew-Zunger (PZ) parameterization of the correlation energy~\cite{PZ}. For all calculations, the $1s^2$ core is replaced by a Troullier-Martins pseudopotential~\cite{TM}. The Kohn-Sham eigenvalues and dipole matrix elements between Kohn-Sham orbitals are computed, and substituted into Eq.~\ref{cavity linewidth} to obtain the bare Kohn-Sham linewidths, $\Gamma_{ij}^{ks}$. These correspond to the real part of the Kohn-Sham-Redfield tensor (imaginary part of the bare Kohn-Sham frequency) of Eq.~\ref{kohn-sham tensor} and are given in the second column of Table~\ref{table: imaginary part}. From the dipole matrix elements between Kohn-Sham orbitals, the bare Kohn-Sham oscillator strengths can be constructed. The real and imaginary parts of the bare Kohn-Sham response function constructed with these quantities are plotted in figures 1-3. 

Next, we perform a standard LR-ATDDFT calculation to obtain the matrix elements of the adiabatic Hartree-exchange-correlation kernel to be inputted in Eq.~\ref{mo casida}. The matrix elements of the kernel obtained are given in table~\ref{table: kernel}. We also include the energies obtained from the standard LR-ATDDFT calculation in column 4 of Table~\ref{table:real part}. For consistency, we use the LDA with modified PZ functional for the exchange-correlation kernel as well.

 Since the operator in Eq.~\ref{mo casida} is explicitly frequency-dependent even when using an adiabatic kernel, Eq.~\ref{open casida} represents a non-linear eigenvalue problem. We solve it using the generalized eigenvalue algorithm presented in~\cite{generalized 1, generalized 2}. The real part of the solutions to Eq.~\ref{open casida} are given in column 3 of Table~\ref{table:real part}, while the imaginary part is given in column 3 of Table~\ref{table: imaginary part}. The real and imaginary parts of the response function obtained are plotted in figures 1-3.

	


\begin{figure}
\begin{center}
\leavevmode
\includegraphics[height=2.5 in, width=3.5 in]{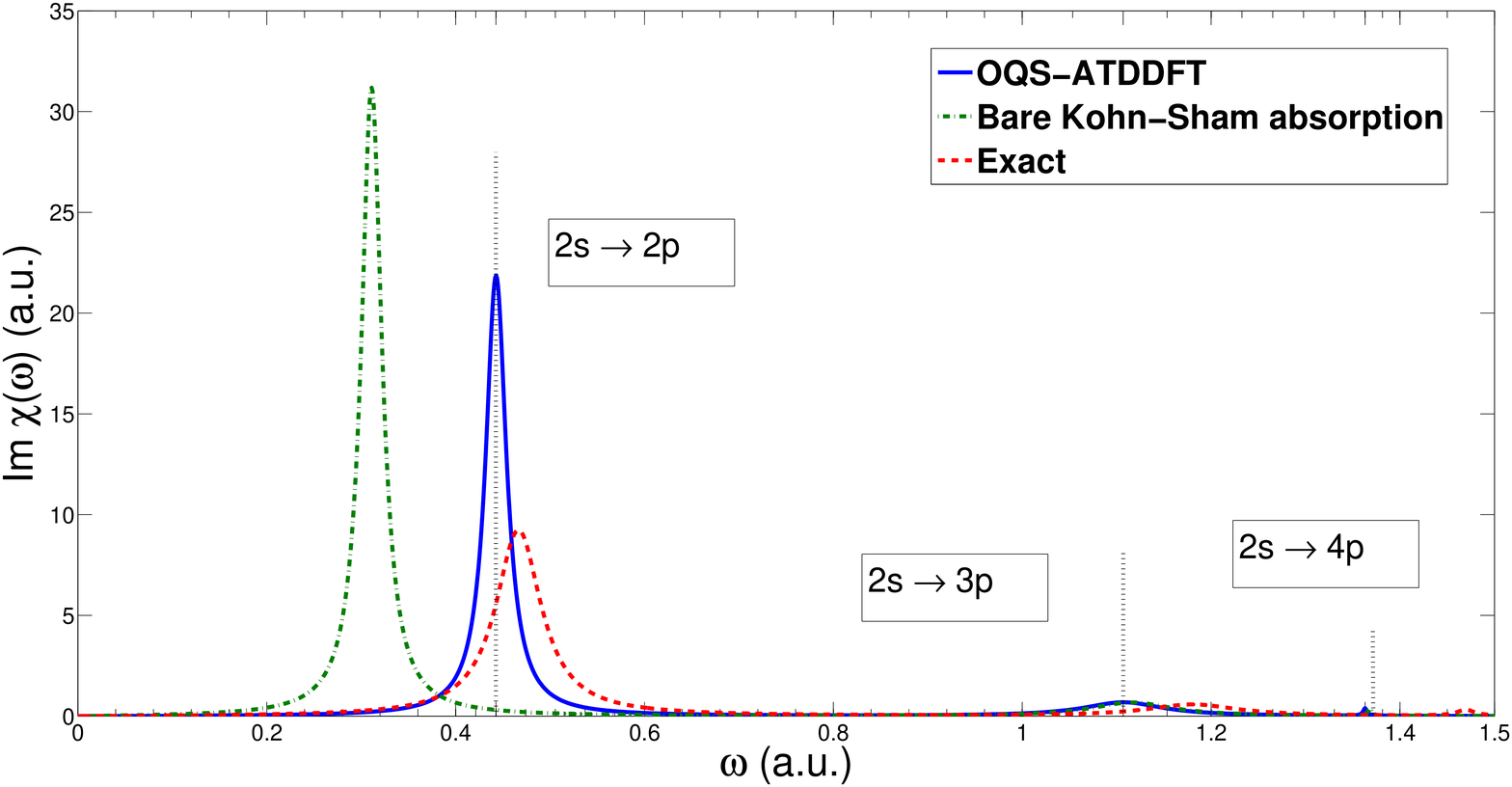}
\end{center}
\caption{Absorption Spectrum of C$^{2+}$ including the 3 lowest dipole allowed transitions. The curves shown are: a) The bare Kohn-Sham spectrum (green-dashed). b) The spectrum obtained by solving Eq.~\ref{open casida} with an adiabatic exchange-correlation kernel (blue-solid). c) The numerically exact spectrum obtained using experimental data (red-dashed). Also shown is the stick spectrum obtained by solving the usual Casida equations for C$^{2+}$ using ALDA (black-dotted).}
\label{fig: Absorption Spectrum}
\end{figure}

\begin{figure}
\begin{center}
\leavevmode
\includegraphics[height=2.5 in, width=3.5 in]{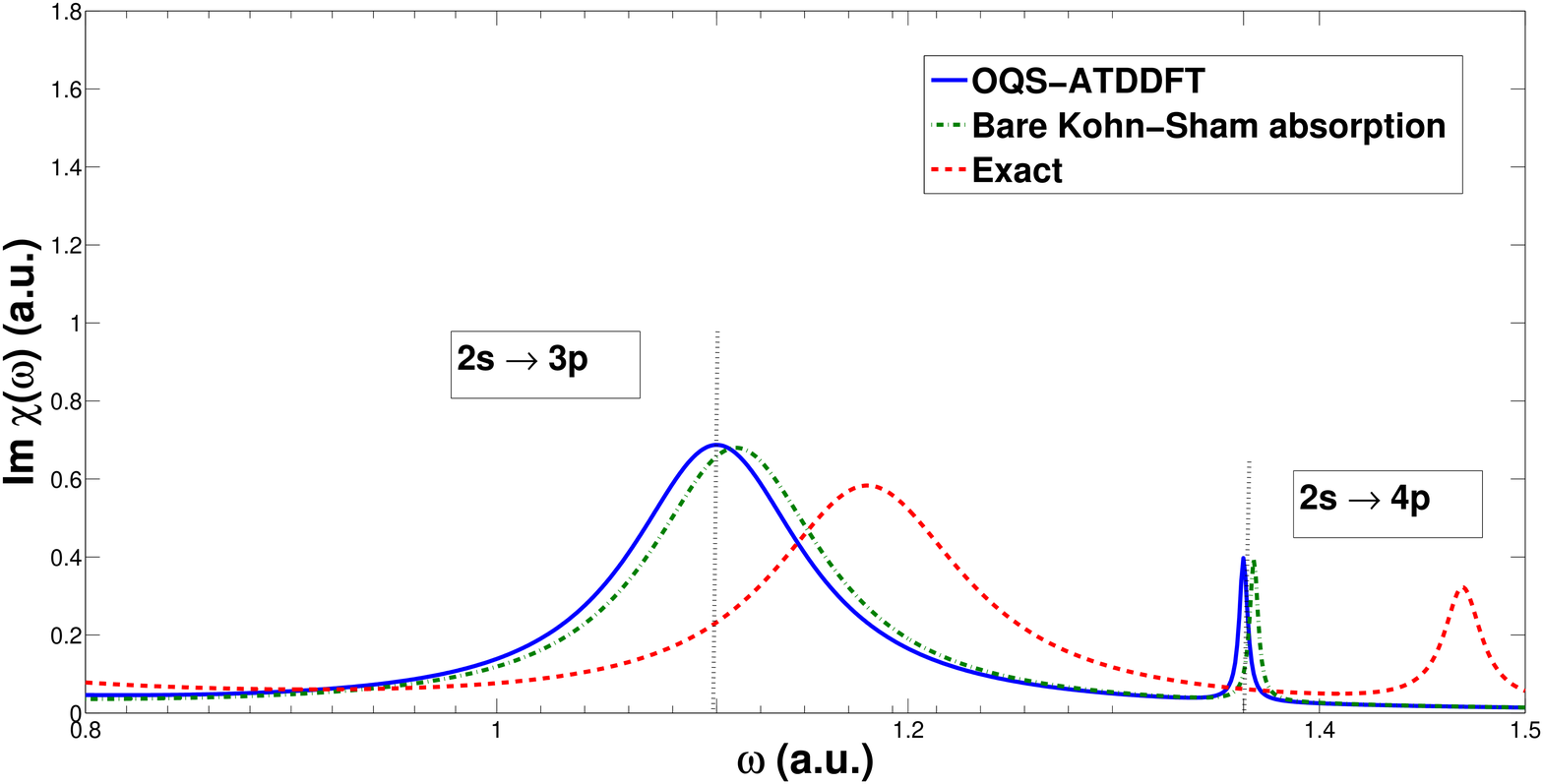}
\end{center}
\caption{Same as Figure 1, but with a close-up view of the $2s \rightarrow 3p$ and $2s \rightarrow 4p$ transitions.}
\label{fig: high lying Absorption Spectrum}
\end{figure}

\begin{figure}
\begin{center}
\leavevmode
\includegraphics[height=2.5 in, width=3.5 in]{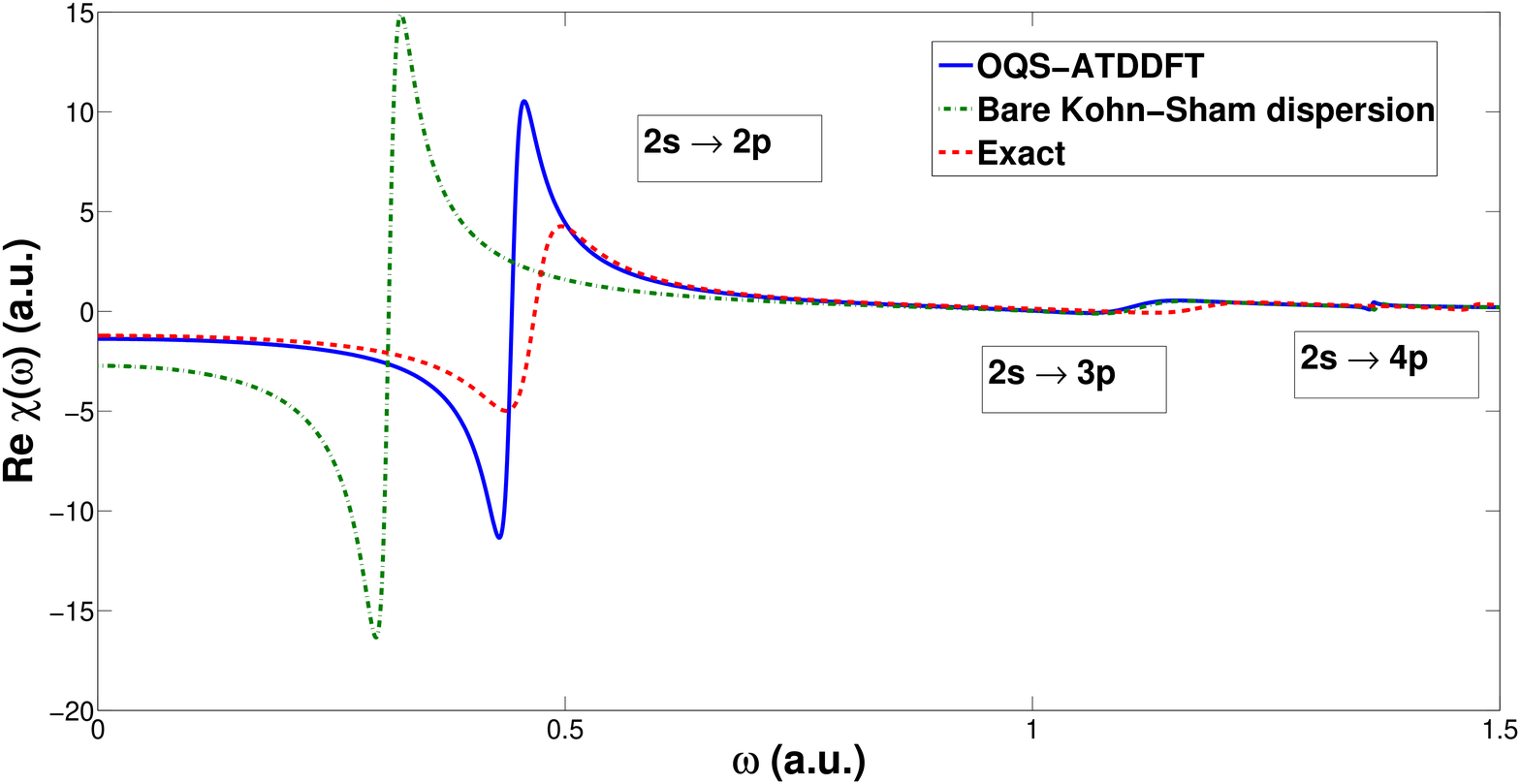}
\end{center}
\caption{Real part of the density-density response function of C$^{2+}$ including the 3 lowest dipole allowed transitions. The effect of the photon bath is to broaden the dispersion over multiple frequencies. The curves shown are: a) The bare Kohn-Sham dispersion (green-dashed). b) The dispersion relation obtained by solving Eq.\ref{open casida} with an adiabatic exchange-correlation kernel (blue-solid). c) The dispersion relation obtained using experimental data (red-dashed).}
\label{fig: Dispersion}
\end{figure}

\begin{table}[ht]
\caption{Real part of the 3 lowest transition frequencies for C$^{2+}$ in an optical resonator in a.u.}
\centering
\begin{tabular}{c c c c c}
\hline\hline
Transition & Bare Kohn-Sham  & OQS-ATDDFT  &ATDDFT&  Exact \\ [0.5ex]
\hline
2s $\rightarrow$ 2p  & 0.311 & 0.443 & 0.443&0.467 \\
2s $\rightarrow$ 3p & 1.116 & 1.107 &1.107 &1.180 \\
2s $\rightarrow$ 4p & 1.368 & 1.361 &1.363 &1.470 \\ 
\hline
\end{tabular}
\label{table:real part}
\end{table}

\begin{table}[ht]
\caption{Imaginary part of the 3 lowest transition frequencies for C$^{2+}$ in an optical resonator in a.u. The last column includes the GL perturbation correction to the 2s $\rightarrow$ 2p transition.}
\centering
\begin{tabular}{c c c c c}
\hline\hline
Transition & Bare Kohn-Sham  &  OQS-ATDDFT  &  Exact & OQS-ATDDFT + GL \\ [0.5ex]
\hline
2s $\rightarrow$ 2p  & 1.329 $\times 10^{-2}$ & 1.331 $\times 10^{-2}$ & 2.932 $\times 10^{-2}$ & 1.805 $\times 10^{-2}$\\
2s $\rightarrow$ 3p & 5.162 $\times 10^{-2}$ & 5.159 $\times 10^{-2}$ & 5.740 $\times 10^{-2}$ &  \\
2s $\rightarrow$ 4p & 2.443 $\times 10^{-3}$ & 2.444 $\times 10^{-3}$ & 1.089 $\times 10^{-2}$ & \\ 
\hline
\end{tabular}
\label{table: imaginary part}
\end{table}

\section{V. Analysis}

\subsection*{Effect of using an adiabatic approximation to $f_{xc}^{open}$}

From Tables~\ref{table:real part} and~\ref{table: imaginary part}, it is clear that using an adiabatic kernel in Eq.~\ref{open casida} gives essentially the same corrections to the real part of the energy as usual LR-ATDDFT, while giving almost no correction to the imaginary part. This means that using an adiabatic kernel in OQS-TDDFT is expected to give the same reliable correction to the location of absorption peaks as usual LR-ATDDFT, while correcting the bare Kohn-Sham linewidths requires an additional frequency-dependent functional. This justifies \textit{a posteriori} our separation of the exchange-correlation kernel in Eq.~\ref{fxc partition} into an adiabatic part and a frequency-dependent part due exclusively to bath effects. 

To understand this situation better, we consider a "small matrix approximation" (SMA), in which a single occupied to unoccupied Kohn-Sham transition, $i \rightarrow j$, is completely isolated from all other transitions~\cite{Appel SMA, Ullrich 2-level, Neepa SMA, Appel DPA}. This is valid when the transition of interest is weakly coupled to all other excitations. In this case, Eq.~\ref{open casida} reduces to

\begin{equation}
\bar{\Omega}(\omega)_{ij,ij} F_{ij} = \omega^2 F_{ij},
\end{equation}

which is equivalent to the polynomial equation

\begin{equation}
\omega^2 + 2 \imath   \Gamma_{ij}^{ks} \omega - \left [(\omega_{ij}^{ks} + \Delta_{ij}^{ks})^2 +({\Gamma_{ij}^{ks}})^2 + 4 (\omega_{ij}^{ks} + \Delta_{ij}^{ks}) K_{ij,ij}(\omega) \right] = 0.
\label{small matrix dissipative}
\end{equation}

If one assumes the adiabatic approximation, the coupling matrix is frequency independent. i.e.

\begin{equation}
K_{ij,ij}(\omega) \approx K_{ij,ij}.
\end{equation}

Eq.~\ref{small matrix dissipative} then reduces to a simple quadratic equation, with solutions given by

\begin{equation}
\omega = - \imath \Gamma_{ij}^{ks} \pm \sqrt{(\omega_{ij}^{ks} + \Delta_{ij}^{ks})^2 + 4 (\omega_{ij}^{ks} + \Delta_{ij}^{ks}) K_{ij,ij}}.
\end{equation}

This shows that the eigenvalues retain their bare Kohn-Sham linewidths.  It is evident from Table~\ref{table: kernel} that in general, the diagonal matrix elements of the kernel are appreciably larger than the off diagonal elements, suggesting that the transitions are weakly coupled. A more rigorous criterion for the validity of the SMA is given in Eq. (11) of~\cite{Appel SMA}. It can be shown that this criterion does in fact hold for the 3 lowest transitions of C$^{2+}$ which we have included in our calculation.

\begin{table}[ht]
\caption{Matrix elements of the adiabatic Hartree-exchange-correlation kernel (Eq.~\ref{coupling matrix}) used in solving Eq.~\ref{open casida}.}
\centering
\begin{tabular}{c c}
\hline\hline
Kernel matrix element & Numerical value  \\ [0.5ex]
\hline
$K_{2s  2p, 2s 2p}$  & 8.035 $\times 10^{-2}$ \\
$K_{2s  3p, 2s 3p}$ &  -4.827 $\times 10^{-3}$  \\
$K_{2s  4p, 2s 4p}$ & -2.528 $\times 10^{-3}$  \\ 
$K_{2s  2p, 2s 3p}$ & 1.031 $\times 10^{-2}$  \\
$K_{2s  2p, 2s 4p}$ & 1.945 $\times 10^{-3}$  \\
$K_{2s  3p, 2s 4p}$ & -4.730 $\times 10^{-4}$  \\
\hline
\end{tabular}
\label{table: kernel}
\end{table}

We also note that for the $2s \rightarrow 3p$ and $2s \rightarrow 4p$ transitions, ATDDFT seems to provide much less reliable results then for the $2s \rightarrow 2p$ transition. This is most likely related to the fact that we have truncated the occupied-unoccupied space to only the 3 lowest transitions. In the following analysis, we will focus on the low-lying $2s \rightarrow 2p$ transition.

\subsection*{Beyond an adiabatic approximation to $f_{xc}^{open}$}

In the previous section, we discussed the effect of approximating $f_{xc}^{open}(\mathbf{r}, \mathbf{r'}; \omega)$ with an adiabatic (frequency independent) kernel. We now investigate what the frequency-dependent contribution $f_{xc}^{bath}(\mathbf{r}, \mathbf{r'}; \omega)$ must be to correct the bare Kohn-Sham linewidths. To formulate our construction rigorously, we examine the difference between the Kohn-Sham-Redfield tensor in Eq.~\ref{kohn-sham tensor} and the interacting Redfield tensor in Eq.~\ref{interacting redfield}. The bath correlation functions in both expressions are the same. The difference lies in the eigenenergies and wavefunctions of the system Hamiltonain used to construct the Redfield tensor. These are Kohn-Sham quantities in the first case and many-body quantities in the latter case. This suggests that the interacting Redfield tensor can be expanded in a G{\"o}rling-Levy (G-L) perturbation series in the electron-electron interaction coupling constant $\alpha$, with the Kohn-Sham-Redfield tensor entering as the zeroth-order term in the series~\cite{Gorling-Levy, Gorling-Levy 2, Gorling Exactexchange, Gorling time-dependent, Gorling excited}:

\begin{equation}
R_{abcd}(\alpha) \approx R_{abcd}(0) + \alpha R_{abcd}^1 + \alpha^2 R_{abcd}^2 + ...
\label{GL exp}
\end{equation}

For linear response from the ground-state as we consider here, we need only construct the quantities $\{ R_{a0a0}(\alpha) \}$.
This is done by first expanding the ground-state and excited-state wavefunctions in a G-L perturbation series in $\alpha$,

\begin{equation}
|a(\alpha) \rangle = \sum_{i=1}^{\infty} \alpha^i |a^i \rangle,
\label{G-L wavefunction}
\end{equation}

as well as the corresponding energies

\begin{equation}
E_a(\alpha) =  \sum_{i=1}^{\infty} \alpha^i E_a^i. 
\label{G-L energies}
\end{equation}

These expansions are then substituted into the general expression for the Redfield tensor to construct an expansion of the Redfield tensor at coupling constant $\alpha$. For $\alpha =1$, we recover the interacting Redfield tensor in Eq.~\ref{interacting redfield}, since Eq.~\ref{G-L wavefunction} and Eq.~\ref{G-L energies} then refer to wavefunctions and energies of the interacting system. For $\alpha=0$, we obtain a Redfield tensor $R_{a0a0}(0)$ written in terms of Kohn-Sham  ground-state and excited Slater determinants, $\{ |a(0) \rangle \}$, and corresponding energies $\{ E_a(0)\}$, which lie on the adiabatic connection with the interacting wavefunctions $\{|a(1) \rangle\}$ and energies $\{E_a(1)\}$. Due to the one-body nature of the system-bath coupling, only matrix elements of the tensor $R_{a0a0}(0)$ containing singly-excited Kohn-Sham Slater determinants are non-zero. The zeroth-order term in the expansion of Eq.~\ref{GL exp} then reduces to,

\begin{equation}
R_{a0a0}(0) = R_{ijij}^{ks}.
\end{equation}

Here, the indices i, j label a pair of Kohn-Sham orbitals, in which an orbital $\phi_i$ occupied in the Kohn-Sham groundstate is replaced by an orbital $\phi_j$ occupied in the singly excited determinant $|a(0) \rangle$. With the G-L expansion of the Redfield tensor well formulated, one can rigorously construct a G-L expansion of  $f_{xc}^{open}(\mathbf{r}, \mathbf{r'}; \omega)$ using the same general procedure outlined in~\cite{Gorling Exactexchange}. 

For the specific example of C$^{2+}$ we consider here with the Lamb shifts neglected, Eq.~\ref{GL exp} amounts to expanding $\Gamma_{a0}$, in a G-L series as

\begin{equation}
\Gamma_{a0}(\alpha) \approx \Gamma_{ij}^{ks} + \alpha \Gamma_{a0}^{1} + ...,
\end{equation}

and determining the corresponding corrections to the bare Kohn-Sham linewidth. This is done explicitly in Appendix B for the $1s^2 2s^2 \rightarrow 1s^2 2s 2p$ transition within the SMA. The first-order correction $\Gamma_{2p, 2s }^{1}$ is found to be

\begin{eqnarray}
&&\Gamma_{2p, 2s }^{1} = -\frac{4}{c^3} M(\epsilon_{2s} -\epsilon_{2p})(\epsilon_{2s} -\epsilon_{2p})^2 \nonumber \\ &\times&  \left[ \int d^3 \mathbf{r} \phi_{2s}(\mathbf{r}) \mathbf{r} \phi_{2p}(\mathbf{r}) \right]^2 \nonumber \\ &\times& \left[ (2s2s| 2s2s) -(2s2s| 2p2p) - (2s2p| 2s2p) \right],
\label{GL correction}
\end{eqnarray}

where $\phi_{2s}$ and $\phi_{2p}$ are ground-state Kohn-Sham orbitals and

\begin{equation}
(ij|kl) = \int d^3 \mathbf{r} \int d^3 \mathbf{r'} \frac{\phi_i(\mathbf{r}) \phi_j(\mathbf{r}) \phi_k(\mathbf{r'}) \phi_l(\mathbf{r'})}{|\mathbf{r} - \mathbf{r'}|}.
\label{coulomb integral}
\end{equation}

The frequency-dependent matrix element of the kernel $f_{xc}^{bath}(\mathbf{r}, \mathbf{r'}; \omega)$ to first-order in G-L perturbation theory is found to be (see Appendix B):

\begin{eqnarray}
K_{2s  2p, 2s 2p}^{bath}(\omega) = -\frac{\imath}{2 (\epsilon_{2s} -\epsilon_{2p})}(\omega + \imath  \Gamma_{2p, 2s }^{ks}) (\Gamma_{2p, 2s }^{1}).
\label{bath kernel}
\end{eqnarray}

To calculate the numerical value of Eq.~\ref{GL correction}, we evaluated the Coulomb integrals of Eq.~\ref{coulomb integral} using the spectral-element code previously presented in \cite{spectral element},
taking the orbitals from Octopus as input.  The Octopus orbitals were
obtained on a uniform mesh with a grid-spacing of 0.1 a.u.  The
spectral-element code projects the orbitals onto a tensorial Chebyshev
basis for the efficient evaluation of Eq.~\ref{coulomb integral}, and we chose all
parameters conservatively such that the main source of error is in the
original representation of the orbitals.

The imaginary part of the frequency of the $2s \rightarrow 2p$ transition obtained by solving Eq.~\ref{open casida} with Eq.~\ref{bath kernel} included is given in column 5 of Table \ref{table: imaginary part}. In Figure 4, the imaginary part of the density-density response function with Eq.~\ref{bath kernel} inculded is plotted in the vicinity of the $2s \rightarrow 2p$ transition. Because we have derived Eq.~\ref{bath kernel} within the SMA, there is no correction to the oscillator strength \cite{Appel SMA, Appel DPA} and the entire effect seen in Figure 4 is due to a change in the linewidth. The contribution from $f_{xc}^{bath}(\mathbf{r}, \mathbf{r'}; \omega)$ provides significant improvement to the linewidth, but is still far from the exact value, suggesting that higher-order corrections are important. 

\begin{figure}
\begin{center}
\leavevmode
\includegraphics[height=2.5in, width=3.5in]{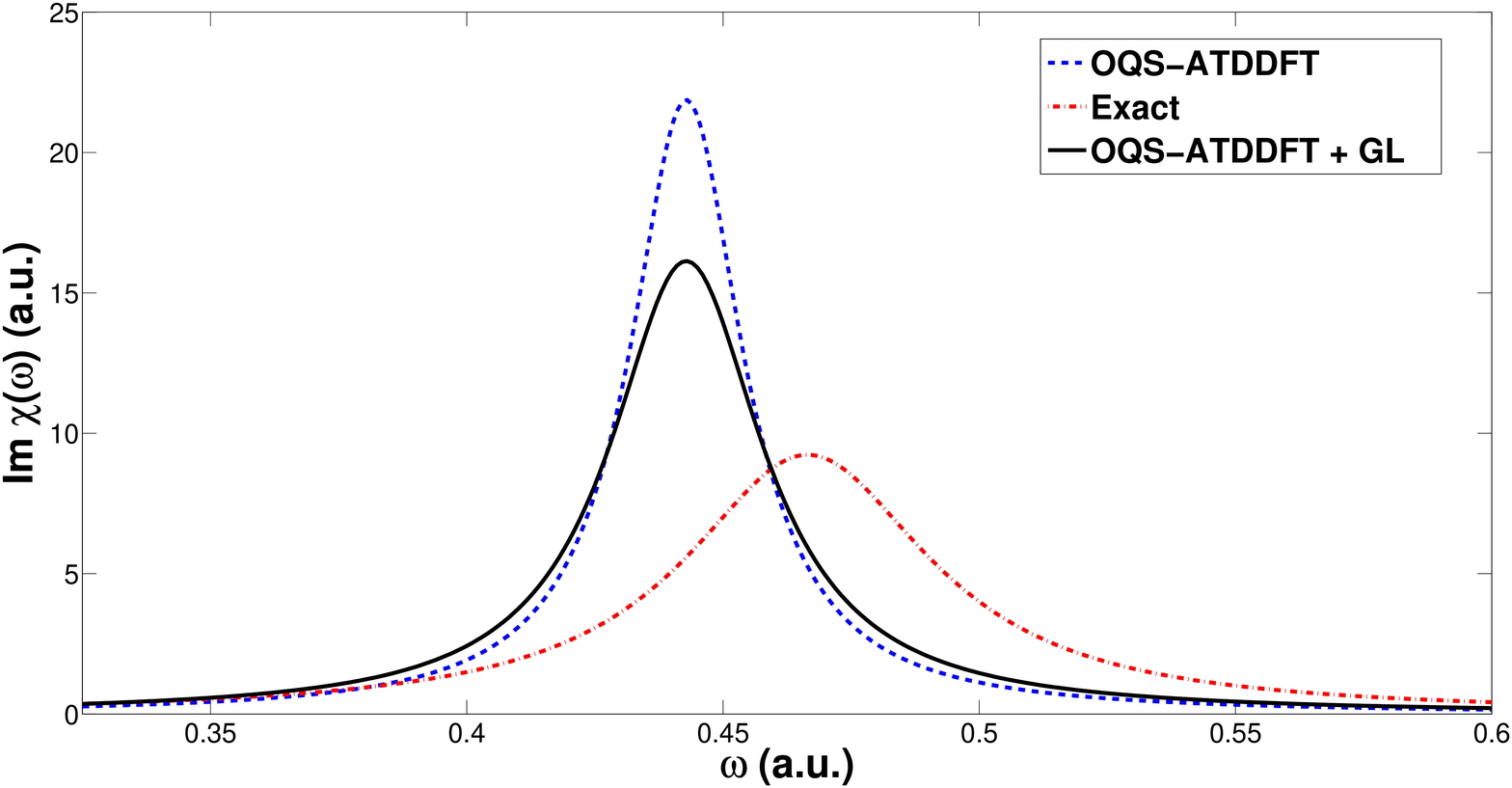}
\end{center}
\caption{Correction to the bare Kohn-Sham linewidth to first-order in GL perturbation theory.}
\label{fig: Gorling_correction}
\end{figure}

The G-L expansion procedure outlined above is expected to provide large corrections to the bare Kohn-Sham linewidths in systems with strong static correlations. In this case, the Kohn-Sham single Slater determinants are a fundamentally poor description of the interacting wavefunctions, and so the bath will interact with the Kohn-Sham system in a very different way than with the interacting system. By including just the first-order correction, one can strongly mix in excited determinants and provide a large correction to the linewidths. However, in the C$^{2+}$ example we have considered, no static correlation is present which might explain why one needs to go to higher orders in the GL expansion to obtain the exact linewidth. Also, we have used the SMA to derive Eq.~\ref{GL correction} and Eq.~\ref{bath kernel}, so we have only allowed mixing in of the first excited Kohn-Sham determinant (see Appendix B).


\section{VI. Conclusion and Outlook}

We have formulated a general framework of LR-TDDFT for many-body open quantum systems, which in principle gives access to environmentally broadened spectra in a strictly \textit{ab initio} manner. Our treatment is most applicable to microscopically derived master equations, in which the system and bath are both treated by starting from an underlying microscopic Hamiltonian. As an example, we analyzed the microscopically derived Redfield master equation for an atom interacting with a photon bath. In this case, the bath correlation functions were very well characterized since they depended only on the free radiation field density of states and the cavity geometry.  In some cases, the bath correlation functions may be more complicated and need to be treated in an approximate way. In the case of a phonon bath, one would need to be able to calculate the phonon density of modes as well as electron-phonon couplings. This could be done by obtaining the vibrational normal modes with a usual DFT calculation, and feeding the couplings and phonon density of states into an open-systems LR-TDDFT calculation for the broadened spectrum. This would be particularly applicable to describing the absorption spectrum of impurity molecules imbedded in a lattice. Similarly, for the case of chromophores in a protein environment, one could simulate the protein using classical molecular dynamics to obtain the spectral density and then feed it into an open-systems LR-TDDFT calculation to compute the absorption spectrum of the chromophore. A similar treatment could be applied to molecules in liquid environments. These directions will be explored in our future work.

The Redfield master equation is Markovian and as a result, the pole structure of the OQS response functions are simple. For Markovian master equations, there are only two parameters characterizing the absorption spectrum;  the location of the peaks and their width.     As a result, the OQS Casida equations have relatively simple frequency-dependence. For non-Markovian master equations, the memory kernel is non-local in time, which corresponds to a frequency-dependent self-energy in the OQS response functions. The resulting lineshapes are asymmetric and may have many non-zero moments. This is expected to give rise to much more complicated frequency dependence in the exchange-correlation kernel, since it must correct all of these moments. A simple step would be to investigate the first-order non-Markovian correction derived from the cumulant expansion of the memory kernel~\cite{Mukamel book}. This will be investigated in future work.

This paper has focused on homogeneous broadening of the spectrum, which arises when the time-scale of the bath is much faster than that of the electrons. This is the case of interest in OQS-TDDFT, since it implies that the bath can induce relaxation and dephasing of the electronic degrees of freedom. In the other limit of inhomogeneous broadening, the bath is static relative to the time-scale of the electrons. In this case, the external potential due to the nuclei is distributed in different configurations due to the local environment, but no relaxation and dephasing takes place. Inhomogeneous broadening can be well captured by performing usual closed LR-TDDFT calculations for different static nuclear configurations and then ensemble averaging. A realistic spectrum usually consists of both broadening mechanisms. This can be well captured by performing an open systems LR-TDDFT calculation for each static nuclear configuration and then ensemble averaging over the different configurations afterwards.

To perform an open-systems LR-TDDFT calculation using modern electronic structure codes, the greatest challenge is probably the implementation of the algorithm in~\cite{generalized 1, generalized 2} for solving the complex and non-linear eigenvalue problem. This algorithm becomes expensive for a large occupied-unoccupied space of Kohn-Sham orbitals and a self consistent procedure becomes more efficient. Work towards implementing these capabilities in a numerical library is currently under investigation.

\section*{Acknowledgements} 

The authors would like to thank Joel Yuen-Zhou, Dmitrij Rappoport, John Parkhill and Neepa Maitra for helpful discussions. D.G.T., R.O.A. and M.A.W. gratefully acknowledge NSF PHY-0835713 for financial support.


\section{Appendix A - Derivation of the Casida equations for the Redfield Master equation}

In this appendix we derive Eq. ~\ref{open casida} and Eq.~\ref{mo casida} for the open-systems Casida equations. In keeping with the original derivation by Casida~\cite{Casida Review}, we introduce second quantized creation and annihilation operators, $a_i^{\dag}$ and $a_i$, for a one-particle orbital basis. In what follows, this will be taken to be a basis of real molecular orbitals from a ground or equilibrium-state Kohn-Sham calculation. An arbitrary one-particle operator $\hat{O}$ is represented in this basis as

\begin{equation}
\hat{O} = \sum_{ij} a_i^{\dag} a_j \langle i| \hat{O} | j \rangle.
\end{equation}

The interacting density-density response function in Eq. (\ref{redfield response}) is given by

\begin{equation}
\chi_{nn}(\mathbf{r}, \mathbf{r'}; \omega)  = \sum_{ijkl} \phi_i(\mathbf{r}) \phi_j(\mathbf{r})\chi_{ijkl}(\omega) \phi_k(\mathbf{r'}) \phi_l(\mathbf{r'}),
\end{equation}

where

\begin{eqnarray}
&& \chi_{ijkl}(\omega) = \sum_{a} [P(E_a)] \sum_b \Bigg\{\frac{\langle a|a_j^{\dagger} a_i|b \rangle \langle b|a_k^{\dagger} a_l|a \rangle}{\omega +\omega_{ab}+\imath R_{abab}} \nonumber \\ &-& \frac{\langle a|a_k^{\dagger} a_l|b \rangle \langle b|a_j^{\dagger} a_i|a \rangle }{\omega -\omega_{ab}+\imath R_{abab}^*} \Bigg\}.
\label{interacting MO representation}
\end{eqnarray}

In this basis, the linear response of the interacting 1-particle reduced density matrix at frequency $\omega$ is

\begin{equation}
\delta \rho_{ij}(\omega) = \sum_{kl} \chi_{ijkl}(\omega) \delta v_{ext}(\omega)_{kl},
\end{equation}

and the linear density response is given by

\begin{equation}
\delta n(\mathbf{r}, \omega) = \sum_{ij} \delta \rho_{ij}(\omega) \phi_j(\mathbf{r}) \phi_i(\mathbf{r}).
\end{equation}

In the Kohn-Sham-Redfield system, the same density response is produced from

\begin{equation}
\delta n(\mathbf{r}, \omega) = \sum_{ij} \delta \gamma_{ij}(\omega) \phi_j(\mathbf{r}) \phi_i(\mathbf{r}),
\end{equation}

where

\begin{equation}
\delta \gamma_{ij}(\omega) = \sum_{kl} \chi_{ijkl}^{ks}(\omega) \delta v_{ks}(\omega)_{kl},
\label{kohn-sham density matrix}
\end{equation}

is the response of the Kohn-Sham-Redfield one-particle reduced density matrix and

\begin{equation}
\chi_{ijkl}^{ks}(\omega) = \delta_{ik} \delta_{jl} \Bigg\{\frac{f_j - f_i}{\omega - \omega_{ij} + \imath r_{jiji}^{ks}} \Bigg\}.
\end{equation}

Eq. (\ref{kohn-sham density matrix}) can then be written as

\begin{equation}
\delta \gamma_{ij}(\omega) = \Bigg\{\frac{f_j - f_i}{\omega - \omega_{ij} + \imath r_{jiji}^{ks}} \Bigg\} (\delta v_{ext}(\omega)_{ij} + \delta v_{h}(\omega)_{ij} + \delta v^{open}_{xc}(\omega)_{ij}).
\label{ks density matrix response}
\end{equation}

We now write

\begin{equation}
\delta v_{h}(\omega)_{ij} + \delta v^{open}_{xc}(\omega)_{ij} = \sum_{kl} K_{ijkl}(\omega) \delta \gamma_{kl}(\omega),
\end{equation}

where

\begin{eqnarray}
&& K_{ijkl}(\omega) = \int d^3 \mathbf{r} \int d^3 \mathbf{r'} \phi_i^*(\mathbf{r}) \phi_{j}^*(\mathbf{r}) \Bigg\{ \frac{1}{|\mathbf{r} - \mathbf{r'}|}  \nonumber \\ && +f_{xc}^{open}[n^{eq}](\mathbf{r}, \mathbf{r'}; \omega) \Bigg\} \phi_k(\mathbf{r'}) \phi_{l}(\mathbf{r'}).
\label{coupling matrix appendix}
\end{eqnarray}

We can then re-write Eq.~\ref{ks density matrix response} as

\begin{equation}
\sum_{kl}^{ f_k \neq f_l} \left [ \delta_{ik} \delta_{jl} \frac{\omega - \omega_{kl} + \imath r_{lklk}^{ks}}{f_l - f_k} - K_{ijkl}(\omega) \right] \delta \gamma_{kl}(\omega) = \delta v_{ext}(\omega)_{ij}.
\end{equation}

We separate the Kohn-Sham-Redfield matrix into real and imaginary parts 

\begin{equation}
r_{klkl}^{ks} = \Gamma_{kl}^{ks} + \imath \Delta_{kl}^{ks},
\end{equation}

and separate particle-hole and hole-particle contributions as in~\cite{Casida Review}

\vspace{2cm}

\begin{eqnarray}
&& \sum_{kl, f_k > f_l} \left [ \delta_{ik} \delta_{jl} \frac{\omega - \omega_{kl} + \Delta_{kl}^{ks}}{f_l - f_k} + \imath \frac{\delta_{ik} \delta_{jl} \Gamma_{kl}^{ks}}{f_l - f_k} - K_{ijkl}(\omega) \right] \delta \gamma_{kl}(\omega) \nonumber \\ &-& \sum_{kl, f_k > f_l} K_{ijlk}(\omega) \delta \gamma_{lk}(\omega) = \delta v_{ext}(\omega)_{ij}
\label{particle-hole}
\end{eqnarray}

\begin{eqnarray}
&& \sum_{kl, f_k > f_l} \left [ \delta_{ik} \delta_{jl} \frac{\omega - \omega_{lk} - \Delta_{kl}^{ks}}{f_k - f_l} + \imath \frac{\delta_{ik} \delta_{jl} \Gamma_{kl}^{ks}}{f_k - f_l} - K_{jilk}(\omega) \right] \delta \gamma_{lk}(\omega) \nonumber \\ &-& \sum_{kl, f_k > f_l} K_{jikl}(\omega)\delta \gamma_{kl}(\omega) = \delta v_{ext}(\omega)_{ji}.
\label{hole-particle}
\end{eqnarray}

We now define the following matrices:

\begin{equation}
A_{ijkl}(\omega) = \delta_{ik} \delta_{jl} \frac{ \omega_{kl} - \Delta_{kl}^{ks}}{f_k - f_l} - K_{ijkl}(\omega),
\label{modified A}
\end{equation}

\begin{equation}
\Gamma_{ijkl} = \frac{\delta_{ik} \delta_{jl} \Gamma_{kl}^{ks}}{f_k - f_l},
\end{equation}

\begin{equation}
B_{ijkl}(\omega) = - K_{ijlk}(\omega),
\end{equation}

and

\begin{equation}
C_{ijkl} = \frac{\delta_{ik} \delta_{jl}}{f_k - f_l}.
\end{equation}

The matrices $B_{ijkl}$ and $C_{ijkl}$ have the same form as in~\cite{Casida Review}. $A_{ijkl}$ is similar, but includes a contribution due to the Lamb shift and $\Gamma_{ijkl}$ is a new term.

We can combine Eq.~\ref{particle-hole} and Eq.~\ref{hole-particle} into a single matrix equation,

\begin{eqnarray}
&& \Bigg\{ \left(\begin{array}{cc}\bar{A} - \imath \bar{\Gamma} & \bar{B} \\\bar{B} & \bar{A} + \imath \bar{\Gamma}\end{array}\right) - \omega \left(\begin{array}{cc}\bar{C} & 0 \\0 & \bar{C}\end{array}\right)  \Bigg\} \left(\begin{array}{c}\vec{\delta \gamma}(\omega) \\\vec{\delta \gamma}^*(\omega) \end{array}\right) \nonumber \\ &=& \left(\begin{array}{c}\vec{\delta v_{ext}}(\omega) \\\vec{\delta v_{ext}}^*(\omega) \end{array}\right),
\end{eqnarray}

or by applying the unitary transformation

\begin{equation}
\frac{1}{\sqrt{2}}\left(\begin{array}{cc}1 & 1 \\-1 & 1\end{array}\right),
\end{equation}

\begin{eqnarray}
&& \left(\begin{array}{cc}\bar{A} + \bar{B} &\imath \bar{\Gamma} + \omega \bar{C} \\ \imath \bar{\Gamma} + \omega \bar{C} & \bar{A} - \bar{B} \end{array}\right)  \left(\begin{array}{c}\Re e ( \vec{\delta \gamma}(\omega)) \\ -\imath \Im m (\vec{\delta \gamma}(\omega) )\end{array}\right) \nonumber \\ &=& \left(\begin{array}{c}\Re e (\vec{\delta v_{ext})}(\omega) \\ -\imath \Im m(\vec{\delta v_{ext}}(\omega) )\end{array}\right).
\label{A-B matrix}
\end{eqnarray}

Without loss of generality we assume the applied perturbation to be real. Since the molecular orbitals are taken to be real, the density response can be calculated from $\Re e ( \vec{\delta \gamma}(\omega))$ alone. From Eq.~\ref{A-B matrix}, we obtain

\begin{eqnarray}
&& \left[(\bar{A} + \bar{B}) - (\imath \bar{\Gamma} + \omega \bar{C})\left[\bar{A} - \bar{B}\right]^{-1}  (\imath \bar{\Gamma} + \omega \bar{C}) \right] \Re e ( \vec{\delta \gamma}(\omega)) \nonumber \\ &=& \Re e (\vec{\delta v_{ext})}(\omega).
\label{uninverted Casida}
\end{eqnarray}

We introduce the matrices

\begin{equation}
\bar{S} = - \bar{C} \left[\bar{A} - \bar{B}\right]^{-1} \bar{C},
\end{equation}

and 

\begin{equation}
\Omega(\omega) = - \bar{S}^{-\frac{1}{2}} \left[\bar{A} + \bar{B} \right] \bar{S}^{-\frac{1}{2}},
\end{equation}

which have the same form as in the usual Casida equations. Eq.~\ref{uninverted Casida} can then be inverted to obtain

\begin{eqnarray}
&& \Re e ( \vec{\delta \gamma}(\omega)) = S^{-\frac{1}{2}} \Big\{ \omega^2 - \Omega(\omega) + \bar{S}^{-\frac{1}{2}} \bar{\Gamma} \left[\bar{A} - \bar{B} \right]^{-1} \bar{\Gamma} \bar{S}^{-\frac{1}{2}} \nonumber \\ &+& \imath \omega (\bar{S}^{-\frac{1}{2}} \bar{\Gamma} \bar{C}^{-1} \bar{S}^{\frac{1}{2}} + \bar{S}^{\frac{1}{2}} \bar{C}^{-1} \bar{\Gamma}  \bar{S}^{-\frac{1}{2}}) \Big\}^{-1} S^{-\frac{1}{2}} \Re e (\vec{\delta v_{ext})}(\omega)
\end{eqnarray}

The poles of the density-density response function are obtained when the operator in brackets vanishes. Defining

\begin{eqnarray}
&& \bar{\Omega}(\omega) =   \Omega(\omega) - \bar{S}^{-\frac{1}{2}} \bar{\Gamma} \left[\bar{A} - \bar{B} \right]^{-1} \bar{\Gamma} \bar{S}^{-\frac{1}{2}}\nonumber \\ &-& \imath \omega (\bar{S}^{-\frac{1}{2}} \bar{\Gamma} \bar{C}^{-1} \bar{S}^{\frac{1}{2}} + \bar{S}^{\frac{1}{2}} \bar{C}^{-1} \bar{\Gamma}  \bar{S}^{-\frac{1}{2}}),
\end{eqnarray}

this is equivalent to solving the pseudo-eigenvalue equation

\begin{equation}
\Big\{\omega^2 - \bar{\Omega}(\omega) \Big\} \vec{F} = 0.
\label{open casida appendix}
\end{equation}

Returning to the basis of Kohn-Sham molecular orbitals, the matrix representation of $\bar{\Omega}(\omega)$ is

\begin{eqnarray}
&& \bar{\Omega}_{ijkl}(\omega) = \delta_{ik} \delta_{jl} \Big\{ (\omega_{lk}^{ks} + \Delta_{kl}^{ks})^2 + (\Gamma_{kl}^{ks})^2 - 2 \imath \omega \Gamma_{kl}^{ks} \Big\} +\nonumber \\ && 4 \sqrt{(f_i - f_j)(\omega_{ji}^{ks} + \Delta_{ij}^{ks})} K_{ijkl}(\omega) \sqrt{(f_k - f_l)(\omega_{lk}^{ks} + \Delta_{kl}^{ks})}.
\label{mo casida appendix}
\end{eqnarray}

\section{Appendix B - First-order G{\"o}rling-Levy perturbation correction to the linewidth of the $2s \rightarrow 2p$ transition of $C^{2+}$}

In this appendix, we derive the first-order correction to the bare Kohn-Sham linewidth in Eq.~\ref{GL correction} as well as the frequency-dependent functional which gives rise to this correction. Our treatment closely parallels that used by G{\"o}rling in deriving the exact-exchange kernel in~\cite{Gorling Exactexchange}. In what follows, the Lamb shifts are neglected. We also make the assumption that the two electrons in the $1s^2$ core are frozen, and their effect on the two valence electrons is taken into account with an effective potential. This is in fact the case for our numerical calculations, since we replace the $1s^2$ core by a pseudopotential. We then effectively have a two-electron singlet in which the interacting ground-state is

\begin{equation}
\psi_{2s^2}(\mathbf{r}, \mathbf{r'}) \equiv \langle \mathbf{r}, \mathbf{r'}|2s^2(\alpha = 1) \rangle, 
\end{equation}

and the first excited state is

\begin{equation}
\psi_{2s2p}(\mathbf{r}, \mathbf{r'}) \equiv \langle \mathbf{r}, \mathbf{r'}|2s2p(\alpha=1) \rangle. 
\end{equation}

We denote the respective energies of these two states by $E_{2s^2} \equiv E_{2s^2}(\alpha=1)$ and $E_{2s2p} \equiv  E_{2s2p}(\alpha=1)$. The corresponding Kohn-Sham ground-state is

\begin{equation}
\Phi_{2s^2}(\mathbf{r}, \mathbf{r'}) \equiv \langle \mathbf{r}, \mathbf{r'}|2s^2(\alpha = 0)   = \phi_{2s}(\mathbf{r}) \phi_{2s}(\mathbf{r'}) , 
\end{equation}

and Kohn-Sham first excited state is

\begin{equation}
\Phi_{2s2p}(\mathbf{r}, \mathbf{r'}) \equiv \langle \mathbf{r}, \mathbf{r'}|2s2p(\alpha = 0)  = \frac{1}{\sqrt{2}} (\phi_{2s}(\mathbf{r}) \phi_{2p}(\mathbf{r'}) + \phi_{2s}(\mathbf{r'}) \phi_{2p}(\mathbf{r})).
\end{equation}

The respective energies are $2 \epsilon_{2s}\equiv E_{2s^2}(\alpha=0)$ and $\epsilon_{2s} + \epsilon_{2p} \equiv  E_{2s2p}(\alpha=0)$.

Our starting point is the linear density response at coupling constant $\alpha$, in the subspace spanned by the $1s^2 2s^2 \rightarrow 1s^2 2s 2p$ transition,

\begin{eqnarray}
&& \delta n(\alpha,  \mathbf{r}, \omega) = \frac{2(E_{2s}(\alpha) - E_{2p}(\alpha))}{(E_{2s}(\alpha) - E_{2p}(\alpha))^2 - (\omega + \imath \Gamma_{2p,2s}(\alpha))^2} \nonumber \\ &\times& \langle 2s^2(\alpha)|\hat{n}(\mathbf{r})|2s2p(\alpha) \rangle \langle 2s2p(\alpha)|\delta \hat{v}(\alpha, \omega)|2s^2(\alpha) \rangle.
\label{density response} 
\end{eqnarray}

For $\alpha = 1$, this expression refers to the density response of the interacting system, while for $\alpha=0$, it describes that of the Kohn-Sham system. In particular, $\delta \hat{v}(\alpha = 1, \omega) = \delta \hat{v}_{ext}(\omega)$ while $\delta \hat{v}(\alpha = 0, \omega) = \delta \hat{v}_{s}(\omega)$. $\delta n(\alpha, \omega, \mathbf{r}) = \delta n(\mathbf{r}, \omega)$ is invariant with respect to the coupling constant $\alpha$. We now expand both sides of Eq.~\ref{density response} in a taylor series in $\alpha$ and equate coefficients of equal powers of $\alpha$ on both sides. At zeroth-order in $\alpha$, we recover the Kohn-Sham response equation

\begin{eqnarray}
&& \delta n(\mathbf{r}, \omega) =  \frac{2 (\epsilon_{2p} - \epsilon_{2s})}{ (\epsilon_{2p} - \epsilon_{2s})^2 - (\omega + \imath \Gamma_{2p,2s}^{ks})^2} \nonumber \\ &\times& \int d^3 \mathbf{r'} \langle 2s^2(0)|\hat{\rho}(\mathbf{r})|2s2p(0) \rangle \langle 2s2p(0)|\hat{\rho}(\mathbf{r'})|2s^2(0) \rangle \delta v_s(\mathbf{r'}, \omega).
\end{eqnarray}

To evaluate the first-order terms, we need the expansions of the wavefunctions, energies and linewidths to first order in $\alpha$. The expansions of the wavefunctions and energies are given by standard GL perturbation theory:

\begin{equation}
|\psi_{2s^2}^1 \rangle = \frac{\langle \psi_{2s2p}(0)|\hat{v}_{ee} - \hat{v}_h - \hat{v}_x | \psi_{2s^2}(0) \rangle}{ \epsilon_{2s} - \epsilon_{2p}} |\psi_{2s 2p}(0) \rangle
\end{equation}

\begin{equation}
|\psi_{2s2p}^1 \rangle = \frac{\langle \psi_{2s^2}(0)|\hat{v}_{ee} - \hat{v}_h - \hat{v}_x | \psi_{2s2p}(0) \rangle}{\epsilon_{2p} - \epsilon_{2s}} |\psi_{2s^2}(0) \rangle
\end{equation}

\begin{equation}
E_{2s^2}^1 = \langle \psi_{2s^2}(0)|\hat{v}_{ee} - \hat{v}_h - \hat{v}_x | \psi_{2s^2}(0) \rangle
\end{equation}

\begin{equation}
E_{2s2p}^1 = \langle \psi_{2s2p}(0)|\hat{v}_{ee} - \hat{v}_h - \hat{v}_x | \psi_{2s2p}(0) \rangle.
\end{equation}

Since the ground-state is a spin singlet, $\hat{v}_x = - \frac{\hat{v}_h}{2}$ and all quantities can be explicitly evaluated. We find

\begin{eqnarray}
&& \langle \mathbf{r}, \mathbf{r'}|\psi_{2s^2}^1 \rangle= \frac{3}{4 (\epsilon_{2s} - \epsilon_{2p})} ( 2s2s|2s2p) \nonumber \\ &\times& \left[ \phi_{2s}(\mathbf{r}) \phi_{2p}(\mathbf{r'}) + \phi_{2s}(\mathbf{r'}) \phi_{2p}(\mathbf{r}) \right ] =0
\label{2s1}
\end{eqnarray}

and

\begin{eqnarray}
&& \langle \mathbf{r}, \mathbf{r'}|\psi_{2s2p}^1 \rangle= \frac{- 3}{2 \sqrt{2} (\epsilon_{2s} - \epsilon_{2p})}  (2s2s|2s2p) \nonumber \\ &\times& \left[ \phi_{2s}(\mathbf{r}) \phi_{2s}(\mathbf{r'})\right ] =0.
\label{2p1}
\end{eqnarray}

i.e. the first-order corrections to the wavefunctions vanish since $(2s2s|2s2p) = 0$.

The first-order G-L correction to the energy is

\begin{eqnarray}
&& \omega_{2s^2, 2s2p}^1 \equiv E_{2s^2}^1 - E_{2s2p}^1 \nonumber \\ &=& (2s2s|2s2s) -(2s2s|2p2p) - (2s2p|2s2p).
\label{omega1}
\end{eqnarray}

We now obtain the first-order correction to the bare Kohn-Sham linewidth. The linewidth for the $2s^2 \rightarrow 2s2p$ transition at coupling constant $\alpha$ is

\begin{eqnarray}
&& \Gamma_{2p, 2s }(\alpha) = -\frac{4}{3}(\frac{1}{c})^3 (\omega_{2s^2, 2s2p}(\alpha))^3 \nonumber \\ &\times& M( \omega_{2s^2, 2s2p}(\alpha))|\langle \psi_{2s^2}(\alpha)| \vec{\hat{\mu}} | \psi_{2s2p}(\alpha) \rangle|^2.
\end{eqnarray}

Expanding in a taylor series in $\alpha$, at zeroth-order we recover the Kohn-Sham linewidth,

\begin{eqnarray}
 && \Gamma_{2p, 2s }(0) \equiv \Gamma_{2p, 2s }^{ks} =  -\frac{4}{3}(\frac{1}{c})^3 (\epsilon_{2s} - \epsilon_{2p})^3 \nonumber \\ &\times& M( \epsilon_{2s} - \epsilon_{2p})|\langle \psi_{2s^2}(0)| \vec{\hat{\mu}} | \psi_{2s2p}(0) \rangle|^2.
\end{eqnarray}

In terms of Kohn-Sham orbitals this is,

\begin{eqnarray}
 && \Gamma_{2p, 2s }(0) \equiv \Gamma_{2p, 2s }^{ks} =  -\frac{4}{3}(\frac{1}{c})^3 (\epsilon_{2s} - \epsilon_{2p})^3 \nonumber \\ &\times& M( \epsilon_{2s} - \epsilon_{2p})|\int d^3 \mathbf{r} \phi_{2s} (\mathbf{r}) \mathbf{r} \phi_{2p}(\mathbf{r})|^2.
\end{eqnarray}

At first-order in $\alpha$ we find

\begin{eqnarray}
&& \Gamma_{2p, 2s }^1 = -\frac{4}{3}(\frac{1}{c})^3 (\epsilon_{2s} - \epsilon_{2p})^3 M( \epsilon_{2s} - \epsilon_{2p})\nonumber \\ &\times& \langle \psi_{2s^2}(0)| \vec{\hat{\mu}} | \psi_{2s2p}(0) \rangle  \left[\langle \psi_{2s^2}(0)| \vec{\hat{\mu}} | \psi_{2s2p}^1 \rangle +\langle \psi_{2s^2}^1| \vec{\hat{\mu}} | \psi_{2s2p}(0) \rangle \right] \nonumber \\ &-&   4 (\frac{1}{c})^3 (\epsilon_{2s} - \epsilon_{2p})^2 M(\epsilon_{2s} - \epsilon_{2p}) \omega_{2s^2, 2s2p}^1\nonumber \\ &\times& |\langle \psi_{2s^2}(0)| \vec{\hat{\mu}} | \psi_{2s2p}(0) \rangle|^2
\label{first-order primitive}
\end{eqnarray}

Using Eqs.~\ref{2s1} - \ref{omega1}, we can now evaluate Eq.~\ref{first-order primitive}. We find that,

\begin{eqnarray}
&&\Gamma_{2p, 2s }^{1} = -4 \left[ \frac{1}{137} \right] ^3 M(\epsilon_{2s} -\epsilon_{2p})(\epsilon_{2s} -\epsilon_{2p})^2 \nonumber \\ &\times&  \left[ \int d^3 \mathbf{r} \phi_{2s}(\mathbf{r}) \mathbf{r} \phi_{2p}(\mathbf{r}) \right]^2 \nonumber \\ &\times& \left[  (2s2s| 2s2s) - (2s2s| 2p2p) - (2s2p| 2s2p) \right].
\label{GL correction appendix}
\end{eqnarray}

which is the correction to the bare Kohn-Sham linewidth we have included in Eq.~\ref{GL correction}. We can now ask what the frequency-dependent exchange-correlation kernel is which gives rise to this correction. This is in some sense a generalization of the exact-exchnage kernel to OQS, in that it is correct to first-order in G-L perturbation theory. However, the form of this functional will now depend on the bath.

To construct the kernel, we now equate coefficients of first-order in $\alpha$ in Eq.~\ref{density response}. The result is

\begin{equation}
\int d^3 \mathbf{r'} \left[\chi_{nn}^{ks}(\omega, \mathbf{r}, \mathbf{r'}) \delta v^1(\omega, \mathbf{r'}) + h_1^{open}(\omega, \mathbf{r}, \mathbf{r'}) \delta v_s(\omega, \mathbf{r'}) \right] =0.
\end{equation}

Here, $ \delta v^1(\omega, \mathbf{r'})$ is the coefficient of the first-order GL expansion of the potential, obtained from

\begin{equation}
\delta v^1(\alpha, \omega, \mathbf{r}) \approx  \delta v_s(\omega, \mathbf{r}) + \alpha \delta v^1(\omega, \mathbf{r}),
\end{equation}

and

\begin{eqnarray}
&& h_1^{open}(\omega, \mathbf{r}, \mathbf{r'}) = \nonumber \\ &&  \Bigg\{ \frac{2 \omega_{2s^2 2s2p}^1}{(\epsilon_{2s} -\epsilon_{2p})^2 - (\omega + \imath \Gamma_{2p, 2s }^{ks})^2} \nonumber \\ &-& \frac{4 (\epsilon_{2s} -\epsilon_{2p}) ((\epsilon_{2s} -\epsilon_{2p}) \omega_{2s^2 2s2p}^1 -\imath \Gamma_{2p, 2s }^{1})((\omega + \imath  \Gamma_{2p, 2s }^{ks})))}{((\epsilon_{2s} -\epsilon_{2p})^2 - (\omega + \imath  \Gamma_{2p, 2s }^{ks})^2)^2}  \Bigg\} \nonumber \\ &\times& \left[2 \phi_{2s}(\mathbf{r})\phi_{2p}(\mathbf{r})\phi_{2s}(\mathbf{r'})\phi_{2p}(\mathbf{r'}) \right].
\end{eqnarray}

We now separate out the part of $h_1^{open}(\omega, \mathbf{r}, \mathbf{r'})$ which contains the correction to the Kohn-Sham linewidth $\Gamma_{2p, 2s }^{1}$:

\begin{eqnarray}
&& h_1^{bath}(\omega, \mathbf{r}, \mathbf{r'}) = \imath \frac{8 (\epsilon_{2s} -\epsilon_{2p})(\Gamma_{2p, 2s }^{1})(\omega + \imath  \Gamma_{2p, 2s }^{ks})}{((\epsilon_{2s} -\epsilon_{2p})^2 - (\omega + \imath  \Gamma_{2p, 2s }^{ks})^2)^2} \nonumber \\ &\times&  \left[\phi_{2s}(\mathbf{r})\phi_{2p}(\mathbf{r})\phi_{2s}(\mathbf{r'})\phi_{2p}(\mathbf{r'}) \right].
\label{bath}
\end{eqnarray}

The functional arising from this correction is then given by:

\begin{eqnarray}
&& f_{x}^{bath}(\omega, \mathbf{r}, \mathbf{r'}) = \int d^3 \mathbf{r''} d^3 \mathbf{r'''} {\chi ^{ks}}^{-1}(\omega, \mathbf{r}, \mathbf{r''})h_1^{bath}(\omega, \mathbf{r''}, \mathbf{r'''})\nonumber \\ &\times& {\chi ^{ks}}^{-1}(\omega, \mathbf{r'''}, \mathbf{r'}).
\label{functional}
\end{eqnarray}

Here, ${\chi^{ks}}^{-1}$ is the inverse of the Kohn-Sham response function

\begin{eqnarray}
&& \chi^{ks}(\omega, \mathbf{r}, \mathbf{r'}) =\frac{4 (\epsilon_{2s} -\epsilon_{2p})}{(\epsilon_{2s} -\epsilon_{2p})^2 - (\omega + \imath \Gamma_{2p, 2s }^{ks}))^2} \nonumber \\ &\times& \left[ \phi_{2s}(\mathbf{r})\phi_{2p}(\mathbf{r})\phi_{2s}(\mathbf{r'})\phi_{2p}(\mathbf{r'}) \right] 
\end{eqnarray}

in the restricted space. For the open-systems Casida equations, we need the matrix element
of $K_{2s  2p, 2s 2p}^{bath}(\omega)$ of Eq.~\ref{functional}, which is given by

\begin{eqnarray}
K_{2s  2p, 2s 2p}^{bath}(\omega) = -\frac{\imath}{2 (\epsilon_{2s} -\epsilon_{2p})}(\omega + \imath  \Gamma_{2p, 2s }^{ks}) (\Gamma_{2p, 2s }^{1}).
\label{bath kernel app}
\end{eqnarray}

This is the matrix element given in Eq.~\ref{bath kernel}. The remaining part of Eq.~\ref{bath} we have not separated out changes only the location of the $2s^2 \rightarrow 2s 2p$ transition and not the width. In the calculations of section V, we have replaced this contribution to the kernel with an adiabatic functional in solving Eq.~\ref{open casida}. 

To understand Eq.~\ref{bath kernel app} better, we consider the SMA equation in the $2s \rightarrow 2p$ subspace:

\begin{eqnarray}
&& \omega^2 + 2 \imath   \Gamma_{2s2p}^{ks} \omega \nonumber - \Big\{(\epsilon_{2s} - \epsilon_{2p})^2 \nonumber \\&+&({\Gamma_{2s2p}^{ks}})^2 + 4 (\epsilon_{2s} - \epsilon_{2p}) K_{2s2p,2s2p}(\omega) \Big\}= 0.
\label{small matrix dissipative app}
\end{eqnarray}

If we separate out the adiabatic and bath parts as

\begin{equation}
K_{2s2p,2s2p}(\omega)  = K_{2s2p,2s2p} -\frac{\imath}{2 (\epsilon_{2s} -\epsilon_{2p})}(\omega + \imath  \Gamma_{2p, 2s }^{ks}) (\Gamma_{2p, 2s }^{1})
\end{equation}

and substitute into Eq.~\ref{small matrix dissipative app} we get:

\begin{eqnarray}
&& \omega^2 + 2 \imath   (\Gamma_{2s2p}^{ks} +\Gamma_{2p, 2s }^{1}) \omega \nonumber - \Big\{(\epsilon_{2s} - \epsilon_{2p})^2 \nonumber \\&+&({\Gamma_{2s2p}^{ks}})^2 + 2\Gamma_{2s2p}^{ks}\Gamma_{2p, 2s }^{1}+ 4 (\epsilon_{2s} - \epsilon_{2p}) K_{2s2p,2s2p} \Big\}\nonumber \\ &=& 0.
\label{small matrix dissipative app 2}
\end{eqnarray}

The solutions are

\begin{eqnarray}
&& \omega = -\imath (\Gamma_{2s2p}^{ks} +\Gamma_{2p, 2s }^{1}) \nonumber \\ &\pm& \sqrt{(\epsilon_{2s} - \epsilon_{2p})^2 - (\Gamma_{2p, 2s }^{1})^2 +4 (\epsilon_{2s} - \epsilon_{2p}) K_{2s2p,2s2p}}
\end{eqnarray}

Since the term $(\Gamma_{2p, 2s }^{1})^2$ is very small relative to the ATDDFT shift, the effect of $K_{2s  2p, 2s 2p}^{bath}(\omega)$ is a simple correction to the bare Kohn-Sham linewidth by $\Gamma_{2p, 2s }^{1}$.

\end{document}